\documentclass[aps,twocolumn,pra,superscriptaddress,amsmath,showpacs,tightenlines,pdflatex,longbibliography]{revtex4-2}
\usepackage{amssymb}
\usepackage{amsmath}
\usepackage{dcolumn}
\usepackage{graphicx}
\usepackage{subfigure}
\usepackage{mathrsfs}
\usepackage{appendix}
\usepackage{graphicx}
\usepackage{booktabs}
\usepackage{color}
\usepackage{bm}
\usepackage{subcaption}
\usepackage{url}
\usepackage[colorlinks]{hyperref}
\hypersetup{%
    plainpages=true,
    breaklinks=true,
    hypertexnames=false,
    pageanchor=true,
    colorlinks=true,
    linkcolor={blue},
    citecolor={blue},
    urlcolor={blue},
    anchorcolor={black}
}
\begin{document}

\title{Nonreciprocal transmission in a cavity-magnon system by rotational Sagnac effect}
\author{Zhe-Qi Yang}
\affiliation{Fujian Key Laboratory of Quantum Information and Quantum Optics, College of Physics and Information Engineering, Fuzhou University, Fuzhou, Fujian 350108, China}
\author{Si-Qi Lin}
\affiliation{Fujian Key Laboratory of Quantum Information and Quantum Optics, College of Physics and Information Engineering, Fuzhou University, Fuzhou, Fujian 350108, China}
\author{Zhi-Rong Zhong}
\email{zhirz@fzu.edu.cn}
\affiliation{Fujian Key Laboratory of Quantum Information and Quantum Optics, College of Physics and Information Engineering, Fuzhou University, Fuzhou, Fujian 350108, China}

\date{\today }

\begin{abstract}
Ultrahigh nonreciprocal transmission has been achieved in a cavity-magnon system, which consists of two whispering gallery modes (WGMs) and a single magnon mode within a magnetic insulator yttrium iron garnet sphere. The nonreciprocal frequency shift induced by the Sagnac effect enables unidirectional transmission of an input field, while suppressing propagation in the opposite direction, thereby facilitating nonreciprocal optical transmission. Within experimentally accessible parameter regimes, the optical isolation ratio can exceed 40 dB, representing the highest isolation ratio reported to date. Furthermore, applying squeezing to the magnon mode further enhances this isolation performance. Additionally, the directionality of light isolation can be reversed simply by modifying the rotation of the WGM cavity. These findings offer promising prospects for developing high-performance, tunable, and compact optical nonreciprocal devices.

\end{abstract}
\maketitle

\section{INTRODUCTION}
Nonreciprocal light propagation, which describes the asymmetric transmission of optical signals between forward and backward directions, plays a critical role in sensitive signal processing and detection, particularly within the precise context of quantum systems \cite{barzanjeh2025nonreciprocityquantumtechnology,sounas2017non,Shui:22,PhysRevA.106.053707}. In stark contrast to reciprocal systems governed by Lorentz reciprocity, nonreciprocal devices break time-reversal symmetry \cite{mariz2022lorentz}, enabling critical functionalities such as optical isolation, unidirectional amplification, and magnetic-free circulators \cite{jalas2013and,doi:10.1126/science.aam6662,10.1063/5.0166231,PhysRevLett.119.147703}.
Conventional nonreciprocal devices typically achieve time-reversal symmetry breaking by employing ferromagnetic compounds and leveraging the Faraday effect in conjunction with strong magnetic fields \cite{Aplet:64,PhysRevA.92.063845,https://doi.org/10.1002/lpor.201900252}, angular momentum biasing in photonic or acoustic crystals \cite{yu2009complete,PhysRevLett.110.093901,doi:10.1126/science.1246957}, nonlinear spatiotemporal modulation \cite{Nazari:14,PhysRevLett.121.123601,doi:10.1126/science.1214383}, and optomechanical coupling \cite{PhysRevA.91.053854,yan2019perfect,PhysRevA.98.063845,shen2016experimental,shen2018reconfigurable,PhysRevA.111.063504}. However, the bulkiness and high cost of ferrite-based systems impede the miniaturization and integration of nonreciprocal devices. Thanks to advances and maturation in micro/nanofabrication techniques, compact and highly tunable cavity-magnon systems provide an attractive platform for achieving nonreciprocal optical transmission.

Magnons, the quantized spin wave excitations associated with spin ordering in magnets, have attracted significant interest across various fields of physics \cite{ZARERAMESHTI20221,YUAN20221,collet2016generation,PhysRevB.97.060405,PhysRevA.71.032317,zhang2015cavity,8pzl-6c5l,PhysRevA.110.063711,Zhang:25}, such as magnon Kerr effect \cite{PhysRevB.94.224410,zhang2019theory}, magnon polaritons, \cite{PhysRevLett.120.057202,zhang2017observation,PhysRevB.91.094423,PhysRevLett.118.217201,PhysRevB.98.174423} and non-Hermitian physics \cite{PhysRevB.99.054404,PhysRevA.103.063708,Zhang:25}. Notably, due to their precisely tunable magnon frequency and coupling strength, long coherence times, and low dissipation \cite{PhysRevLett.113.083603,doi:10.1126/science.aaa3693,PhysRevLett.104.077202,PhysRevLett.113.156401,PhysRevB.93.144420,PhysRevLett.121.137203,yao2017cooperative}, cavity-magnon systems have great potential for applications in the field of quantum technology \cite{Lachance-Quirion_2019}, for example, quantum sensing \cite{PhysRevB.109.L041301,Flower_2019,PhysRevA.103.052419}, quantum transducers \cite{PhysRevLett.116.223601,PhysRevLett.117.133602,PhysRevB.93.174427}, quantum memory \cite{zhang2015magnon,PhysRevLett.127.183202}, etc. In recent years, cavity-magnon systems have been increasingly explored for achieving nonreciprocal transmission via various approaches \cite{PhysRevApplied.12.034001,Ren:21,Kong:21,PhysRevLett.123.127202,MING2024107915,PhysRevA.110.043704}. However, current approaches to nonreciprocal optical transmission often face challenges such as low isolation ratios and transmission efficiencies, leading to systems with reduced performance, decreased reliability, and increased susceptibility to interference. For example, in Ref. \cite{PhysRevApplied.12.034001}, nonreciprocal transmission has been achieved by utilizing the magnon Kerr nonlinearity, reaching a maximum isolation ratio of 21 dB with a forward transmission coefficient of 0.06. In another work, Ref. \cite{Kong:21} has demonstrated nonreciprocity through quantum interference between different transmission paths in a three-mode cavity magnonics system, attaining a maximum isolation of 34 dB and a forward transmission coefficient of 0.13. Therefore, developing a method to enhance nonreciprocal transmission isolation while maintaining flexibility and ease of control remains a significant challenge.

In the present work, we propose a theoretical scheme to  achieve ultrahigh nonreciprocal transmission in a cavity-magnon system by rotational Sagnac effect. In the scheme, the setup consists of two whispering gallery modes (WGMs) and a single magnon mode within a magnetic insulator yttrium iron garnet (YIG) sphere. When a WGM cavity rotates, the Sagnac effect causes its clockwise (CW) and counterclockwise (CCW) propagating modes to experience opposite Fizeau frequency shifts. This nonreciprocal frequency shift causes the input light field to be transmitted in one direction while being absorbed in the opposite direction, thereby achieving nonreciprocal optical transmission. Further analysis demonstrates that, using feasible experimental parameters, the system achieves an isolation ratio readily exceeding 40 dB, representing a significant improvement over previous work. Additionally, applying squeezing to the magnon mode provides an additional method for enhancement. Crucially, the direction of nonreciprocal transmission can be straightforwardly reversed by changing the WGM cavity's rotation direction, eliminating the need to adjust other system parameters and thereby offering significant practical advantages for controlling nonreciprocal transmission. Our work presents a promising avenue toward high-isolation, on-chip integrated optical nonreciprocal devices, with important implications for optical communication.

\section{THEORETICAL MODEL}
\begin{figure}
\centering
\includegraphics[width=3.3in]{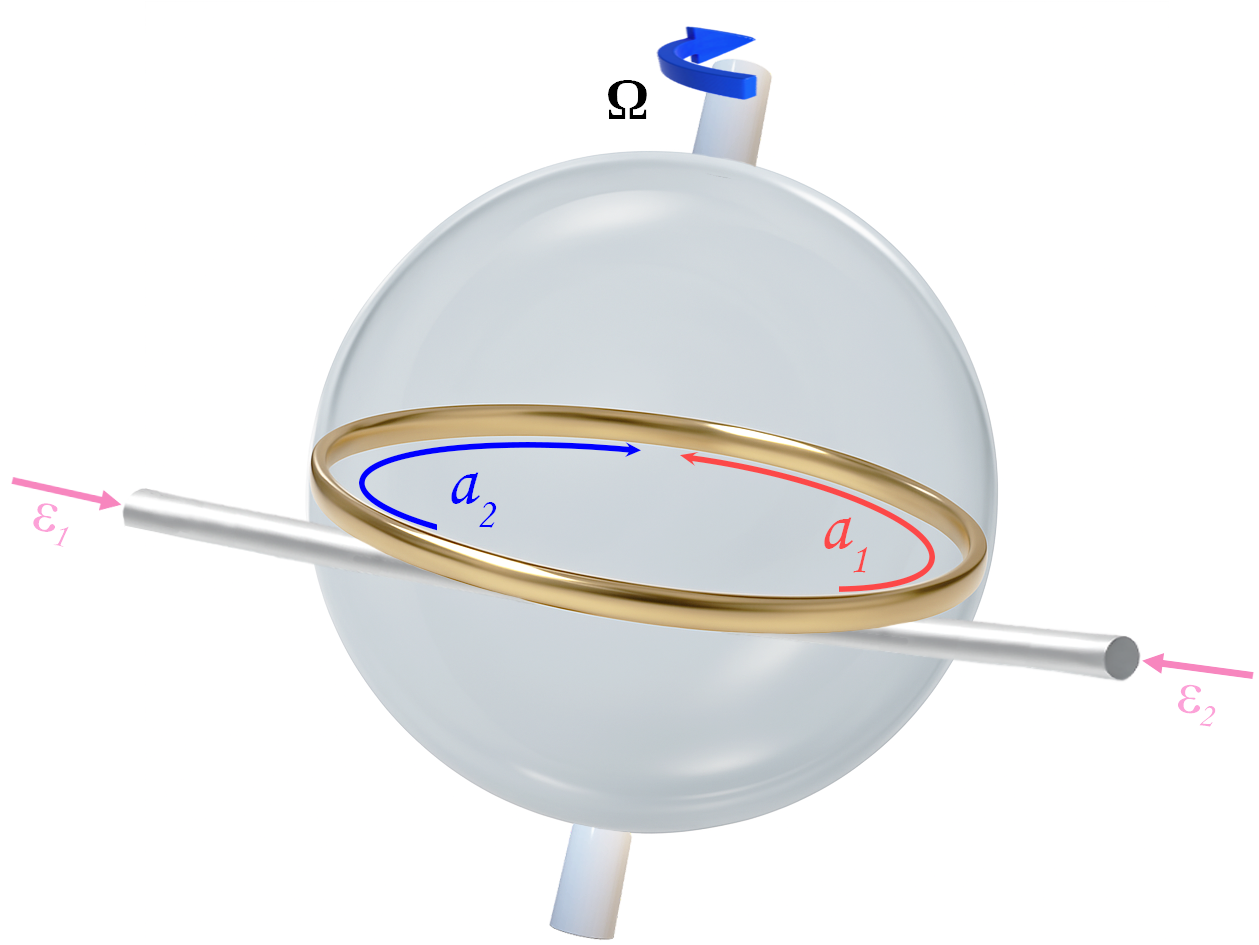}
 \caption{Sketch of the system. A WGM cavity rotates CW with an angular velocity $\Omega$. This cavity supports two optical modes propagating in CW and CCW directions, along with a magnetostatic mode within a YIG sphere. }
 \label{fig:fig1}
\end{figure}
As illustrated in Fig. \ref{fig:fig1}, we consider two optical WGMs and a magnetostatic mode supported by a YIG sphere. Coupling between the cavity mode and the magnon mode is achieved through three-wave mixing \cite{Ren:21,Xie:23}. When a cavity rotates, the resonant frequencies of its CCW and CW propagating modes undergo a Fizeau shift \cite{Maayani2018}:
\begin{eqnarray}
\Delta_F = \pm\Omega\frac{nr\omega_0}{c}(1-\frac{1}{n^2}-\frac{\lambda dn}{n d\lambda}), \label{eq1}
\end{eqnarray}
where $\omega_0$ is the resonance frequency of a nonspinning resonator. The parameter $\Omega$ is the angular velocity of the spinning resonator, $n$ is the refractive index, $r$ is the radius of the resonator, and $dn/d \lambda$ describes the dispersion of the medium, while $c$ and $\lambda$ represent the speed of light and the wavelength of the photon in a vacuum, respectively. When the WGM cavity rotates CW, the Fizeau shifts are $\Delta_F > 0$ for the CCW mode and $\Delta_F < 0$ for the CW mode.

The total Hamiltonian of the system can be written as ($\hbar=1$) \cite{Ren:21,Xie:23}
\begin{eqnarray}
 H =H_0+H_{\rm{int}}+H_{\rm{d}}+H_{sq},
\end{eqnarray}
where
\begin{align}
H_0&= \sum_{j=1,2}(\omega_{0}+\Delta_F)a_j^{\dagger}a_j+\omega _{m}m^{\dagger}m, \label{eq2}\\
H_{\rm{int}}&= \sum_{j=1,2}g_j(a_j^{\dag}m + a_j m^{\dag}), \label{eq3}\\
H_{d}&= \sum_{j=1,2}i\sqrt{\eta_j \kappa_j}\varepsilon_j(a_j^{\dag}e^{-i\omega_{pj} t} -a_je^{i\omega_{pj} t}) \nonumber \\
&+ i\sqrt{\eta_3 \gamma_m}\varepsilon_3(m^{\dag}e^{-i\omega_{p3} t} -me^{i\omega_3 t}),\\
H_{\textrm{sq}}&=\frac{E}{2}(m^2 e^{i\omega_d t} +m^{\dagger2}e^{-i\omega_d t}).
\end{align}
Here, $H_0$ is the free Hamiltonian of two WGMs and a magnon mode. $a_j$ and $a_j^{\dagger}$ are the annihilation and creation operators of the cavity modes, respectively. $\Delta_F > 0$ for the $a_1$ mode and $\Delta_F < 0$ for the $a_2$ mode. The boson operators $m$ and $m^{\dagger}$ represent the annihilation and creation operators for the magnon mode with frequency $\omega _{m}$, which can be tuned by an external magnetic field $H$ via $\omega _{m}=\gamma H$, where $\gamma/2\pi=28$  $\rm{GHz/T}$ is the gyromagnetic ratio. $H_{\rm{int}}$ represents the interaction between the cavity modes and the magnon mode, where $g_j$ is the coupling strength. $H_d$ is the laser-driven term, with an amplitude $\varepsilon_j=\sqrt{P_j/\hbar \omega_{pj}}$ ($j=1,2,3$), where $P_j$ and $\omega_{pj}$ are the amplitude and frequency of the input field, respectively. $\kappa_j$ ($j=1,2$) represents the total dissipation rate of the cavity mode, which includes the intrinsic dissipation $\kappa_{j,0}$ and the external dissipation $\kappa_{j,e}$ \cite{RevModPhys.86.1391}. $\gamma_m$ is the total loss rate of the magnon mode. The coupling parameter $\eta_j=\kappa_{j,e}/\kappa_j$ is the ratio of external loss rate to the total loss rate, which describes the coupling between the driving field and the cavity mode. Similarly, $\eta_3$ is the magnon coupling parameter. $H_{\textrm{sq}}$ represents magnon squeezing via parametric driving \cite{PhysRevA.99.021801,PhysRevA.108.063703,Ding:22,PhysRevA.111.053707,PhysRevA.100.062501}, utilizing a pump field with frequency $\omega_d$ and amplitude $E$.

Considering $\omega_{p1}=\omega_{p2}=\omega_{p3}=\omega_d /2$, we apply a unitary transformation $V=\exp \left[ -i\frac{\omega_d}{2} (a_1^{\dagger}a_1+ a_2^{\dagger}a_2+m^{\dagger}m)t\right]$ to transform the Hamiltonian $H$ to $ H_{1}=V^{\dagger}HV+i \frac{\partial V^{\dagger}}{\partial t}V$. Thus, we obtain
a Hamiltonian
\begin{align}
H_1&= \sum_{j=1,2}[\Delta_ja_j^{\dagger}a_j+g_j(a_j^{\dag}m + a_j m^{\dag})] \nonumber \\
&+ \sum_{j=1,2}i\sqrt{\eta_j \kappa_j}\varepsilon_j(a_j^{\dag} -a_j)+ i\sqrt{\eta_3 \gamma_m}\varepsilon_3(m^{\dag} -m) \nonumber \\
&+ \frac{E}{2}(m^2  +m^{\dagger2}) +\Delta_m m^{\dagger}m,
\end{align}
where $\Delta_1=\Delta+\Delta_F$ ($\Delta_F > 0$), $\Delta_2=\Delta+\Delta_F$ ($\Delta_F < 0$), $\Delta=\omega_{0}-\omega_d/2$, and $\Delta_m=\omega_m-\omega_d/2$. By performing a Bogoliubov transformation
\begin{align}
m_s=\cosh(G) m + \sinh(G) m^{\dag}
\end{align}
on the magnon mode and applying the rotating-wave approximation, we obtain the effective Hamiltonian
\begin{align}
H_{\textrm{eff}}&= \sum_{j=1,2}[\Delta_ja_j^{\dagger}a_j+g_j^{'}(a_j^{\dag}m_s + a_j m_s^{\dag})] \nonumber \\
&+ \sum_{j=1,2}i\sqrt{\eta_j \kappa_j}\varepsilon_j(a_j^{\dag} -a_j)+ i\sqrt{\eta_3 \gamma_m}\varepsilon_3^{'}(m_s^{\dag} -m_s) \nonumber \\
&+ \omega_s m_s^{\dagger}m_s,
\end{align}
where $G=\frac{1}{4}\ln(\frac{\Delta_m+E}{\Delta_m-E})$, $g_j^{'}=g_j \cosh(2G)$ ($j=1,2$), $\varepsilon_3^{'}=\varepsilon_3 e^{-G}$, and $\omega_s=\sqrt{\Delta_m^2-E^2}$.

In the long-time limit, we consider the expectation values of the operators, i.e., their steady-state solutions. We set $A_1=\langle a_1\rangle$, $A_2=\langle a_2\rangle$, and $M=\langle m_s\rangle$. The quantum noise terms are neglected as their expectation values are zero. The evolution of the entire system can then be described by the quantum Langevin equations
\begin{align}
&-i\Delta_1 A_{1}-ig_1^{'} M-\frac{\kappa_1}{2}A_{1}+ \sqrt{\eta_1\kappa_1}\varepsilon_1=0,\nonumber \\
&-i\Delta_2 A_{2}-ig_2^{'} M-\frac{\kappa_2}{2}A_{2}+ \sqrt{\eta_2\kappa_2}\varepsilon_2=0,\nonumber \\
&-i\omega_s M-ig_1^{'}A_{1}-ig_2^{'}A_{2}-\frac{\gamma_m}{2}M+ \sqrt{\eta_3\gamma_m}\varepsilon_3^{'}=0. \label{eq10}
\end{align}
To account for nonreciprocal effects, we consider two input conditions: driving field input from the left side of the cavity, corresponding to $\varepsilon_1 \neq 0$ and $\varepsilon_2=0$; and from the right side, corresponding to $\varepsilon_1 = 0$ and $\varepsilon_2\neq0$. In fact, we only need to analyze the nonreciprocal effect for one direction of light transmission. The effect in the reverse direction is equivalent to that obtained by reversing the rotation of the resonator.

When $\varepsilon_1 \neq 0$ and $\varepsilon_2=0$, the steady-state solution of Eq. (\ref{eq10}) can be obtained as follows:
\begin{align}
A_1&= \frac{F_1}{D_1}-\frac{ig_1^{'}(D_1D_2F_3-ig_1^{'}D_2F_1)}{D_1(D_1D_2D_m+D_2{g_1^{'}}^2+D_1{g_2^{'}}^2)}, \label{eq11} \\
A_2&= \frac{-ig_2^{'}(D_1F_3-ig_1^{'}F_1)}{D_1D_2D_m+D_2{g_1^{'}}^2+D_1{g_2^{'}}^2}, \label{eq12}
\end{align}
where $D_1= i \Delta_1 + \frac{\kappa_1}{2}$, $D_2=i\Delta_2 + \frac{\kappa_2}{2}$, $D_m=i\omega_s + \frac{\gamma_m}{2}$, $F_1=\sqrt{\eta_1\kappa_1}\varepsilon_1$, $F_2=\sqrt{\eta_2\kappa_2}\varepsilon_2$, and $F_3=\sqrt{\eta_3\gamma_m}\varepsilon_3^{'}$. Substituting Eq. (\ref{eq12}) into the standard input-output relation, the output fields $A_{2,\mathrm{out}}$ can be written as follows:
\begin{align}
A_{2,\mathrm{out}}&= \sqrt{\eta_2 \kappa_2}A_2 \nonumber \\
&=\frac{-ig_2^{'}(D_1F_3-ig_1^{'}F_1)\sqrt{\eta_2 \kappa_2}}{D_1D_2D_m+D_2{g_1^{'}}^2+D_1{g_2^{'}}^2}. \label{eq13}
\end{align}
When $\varepsilon_1 = 0$ and $\varepsilon_2 \neq 0$, using the same method, the expression for the output field $A_{1,\mathrm{out}}$ can be obtained as
\begin{align}
A_{1,\mathrm{out}}&= \sqrt{\eta_1 \kappa_1}A_1 \nonumber \\
&=\frac{-ig_1^{'}(D_2F_3-ig_2^{'}F_2)\sqrt{\eta_1 \kappa_1}}{D_1D_2D_m+D_2{g_1^{'}}^2+D_1{g_2^{'}}^2}.\label{eq14}
\end{align}

Subsequently, we define $T_{12}$ and $T_{21}$ as the forward and backward transmission coefficients, respectively, with
\begin{align}
T_{12}&= \left| \frac{A_{1,\mathrm{out}}}{\varepsilon_2} \right|,\nonumber \\
T_{21}&= \left|\frac{A_{2,\mathrm{out}}}{\varepsilon_1}  \right|.
\end{align}
From Eqs. (\ref{eq13}) and (\ref{eq14}), it is clear that when the system parameters (i.e., $g_j^{'}$, $\eta_j$, $\kappa_j$, and $\varepsilon_j$) are fixed, the factor influencing the efficiency of nonreciprocal transmission is the difference between $D_1$ and $D_2$. This difference, in turn, is related to the angular velocity $\Omega$ of the resonator's rotation. Therefore, by adjusting $\Omega$, we can achieve tunable optical nonreciprocal transmission. To more intuitively assess the efficiency of the nonreciprocal transmission, we define the isolation ratio by the following expression \cite{Kong:21,PhysRevApplied.12.034001}:
\begin{align}
I= 10 \left|\log_{10} \frac{|A_{1,\mathrm{out}}|^2}{|A_{2,\mathrm{out}}|^2}\right|.\label{eq16}
\end{align}
When $A_{2,\mathrm{out}}$ = $A_{1,\mathrm{out}}$, the isolation $I = 0$, corresponding to the case of reciprocal transmission. A larger value of $I$ indicates higher nonreciprocal transmission efficiency. Therefore, we will subsequently focus on the influence of system parameters on nonreciprocal transmission.
\begin{figure}[ht]
    \centering
    \begin{minipage}[t]{0.8\columnwidth}
        \centering
        \includegraphics[width=\textwidth]{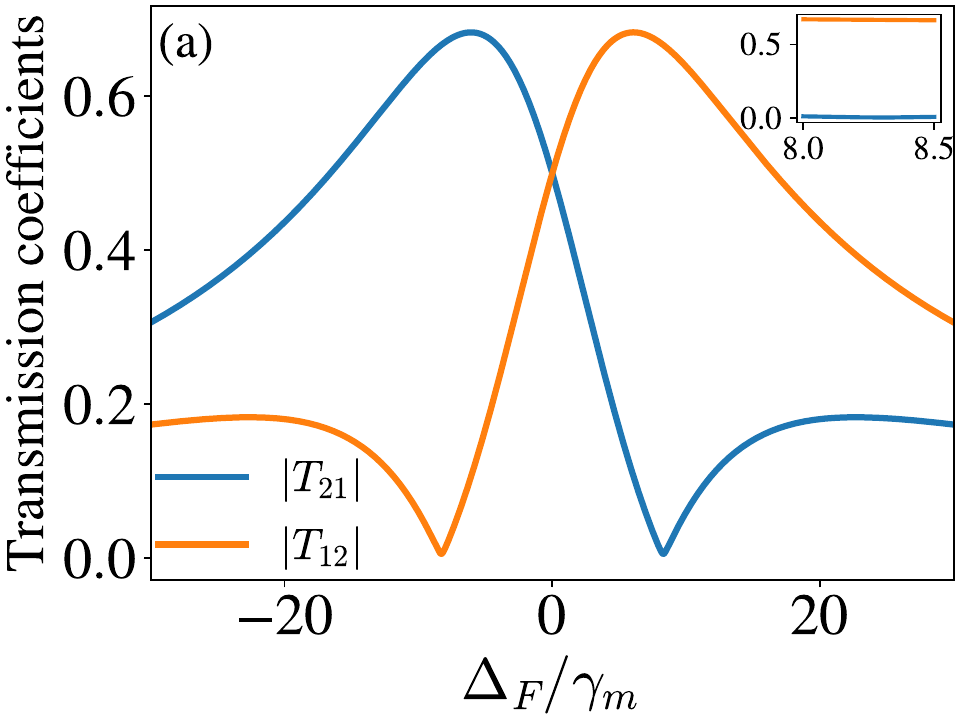}

        \label{fig:2a}
    \end{minipage}\hfill
    \begin{minipage}[t]{0.8\columnwidth}
        \centering
        \includegraphics[width=\textwidth]{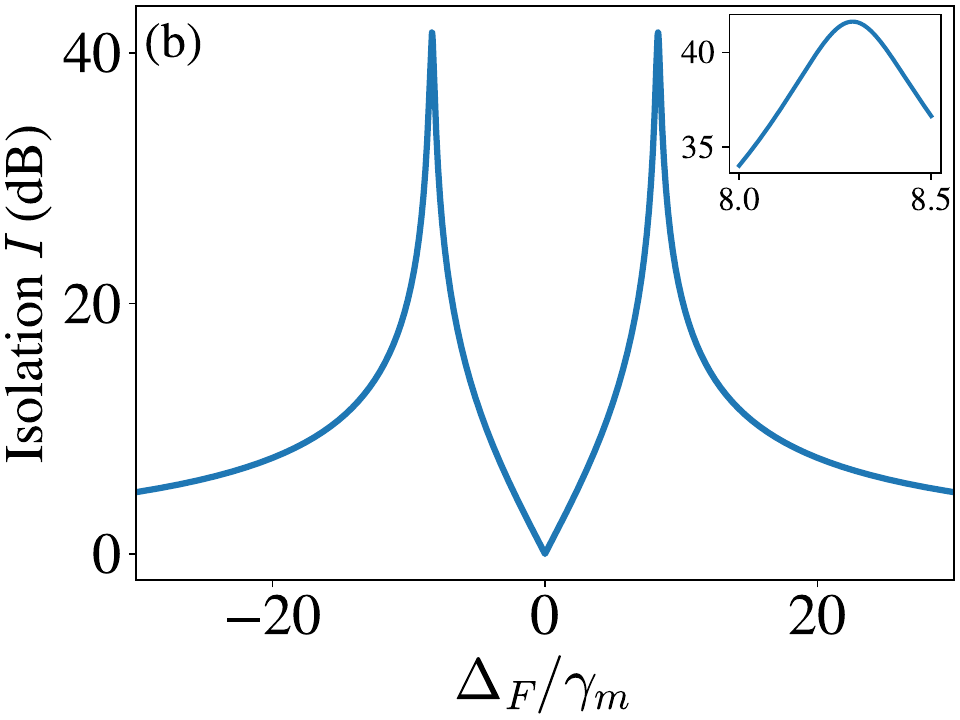}

        \label{fig:2b}
    \end{minipage}
     \caption{(a) Transmission coefficients as a function of $\Delta_F/\gamma_m$. (b) The isolation ratio as a function of $\Delta_F/\gamma_m$. The parameters are $\Delta=0$, $g_0/2\pi$=41 MHz, $G=0.5$, $\kappa /2\pi$ = 1.1 MHz, $\gamma_m/2\pi$=4 MHz, $\eta=0.5$, $P_1=P_2=P_3=100$ mW, $\omega_m/2\pi$=10.1 GHz, which are typical experimentally utilized parameters \cite{ZARERAMESHTI20221,Maayani2018,PhysRevResearch.1.023021,righini2011whispering}.}
    \label{fig:fig2}
\end{figure}
\section{NONRECIPROCAL OPTICAL TRANSMISSION}
In what follows, we will examine the effect of system parameters on the efficiency of nonreciprocal transmission. For simplicity, we set $\kappa_1 = \kappa_2 = \kappa$, $g_1^{'}=g_2^{'}=g=g_0 \cosh(2G)$, $\eta_1=\eta_2=\eta_3=\eta$, and $\varepsilon_1=\varepsilon_2=\varepsilon_3^{'}$. As $\Delta_F$ is tunable, we chiefly focus on how the nonreciprocal transmission efficiency changes with $\Delta_F$ under different system conditions. To facilitate the subsequent analysis, we define $R = |\frac{A_{1,\mathrm{out}}}{A_{2,\mathrm{out}}}|^2$ as the intensity ratio of the forward output field to the backward output field. Differentiating $R$ with respect to $\Delta_F$ and setting $\frac{dR}{d(\Delta_F)}=0$, we obtain
\begin{align}
\Delta_{F_{1,2}} &= \pm \sqrt{\frac{\kappa^2}{4} + (\Delta-g\sqrt{\frac{\kappa}{\gamma_m}})^2},  \label{eq17}  \\
R(\Delta_{F_{1}}) &= [R(\Delta_{F_{2}})]^{-1} \label{eq18} \nonumber \\
&= \frac{\sqrt{\frac{\kappa^2}{4} + \left(\Delta - g\sqrt{\frac{\kappa}{\gamma_m}}\right)^2} - \left(\Delta - g\sqrt{\frac{\kappa}{\gamma_m}}\right)}{\sqrt{\frac{\kappa^2}{4} + \left(\Delta - g\sqrt{\frac{\kappa}{\gamma_m}}\right)^2} + \left(\Delta - g\sqrt{\frac{\kappa}{\gamma_m}}\right)},
\end{align}
where $\Delta_{F_{1}}$ and $\Delta_{F_{2}}$ are the two points at which $R$ exhibits an extremum. Please note that at this point, $\Delta \neq g\sqrt{\frac{\kappa}{\gamma_m}}$. We will discuss the case where $\Delta=g\sqrt{\frac{\kappa}{\gamma_m}}$ later in the text. As is evident from Eq. (\ref{eq16}), the values $R(\Delta_{F_{1}})$ and $R(\Delta_{F_{2}})$ result in the same isolation ratio $I$. This implies that $I$ is maximized to the same value at both points, $\Delta_{F_{1}}$ and $\Delta_{F_{2}}$. It should be emphasized that, experimentally, $\Omega/2\pi$ is typically on the order of kilohertz \cite{Maayani2018}. This implies that $\Delta_F$ lie approximately within the interval (-65, 65) MHz (see Sec. \ref{IV} for details). Thus, all subsequent analyses are performed within this interval. The condition $\Delta_F < 0$ in this context corresponds to the scenario of the resonator rotating in the opposite direction.
\begin{figure}[ht]
    \centering
    \begin{minipage}[t]{0.88\columnwidth}
        \centering
        \includegraphics[width=\textwidth]{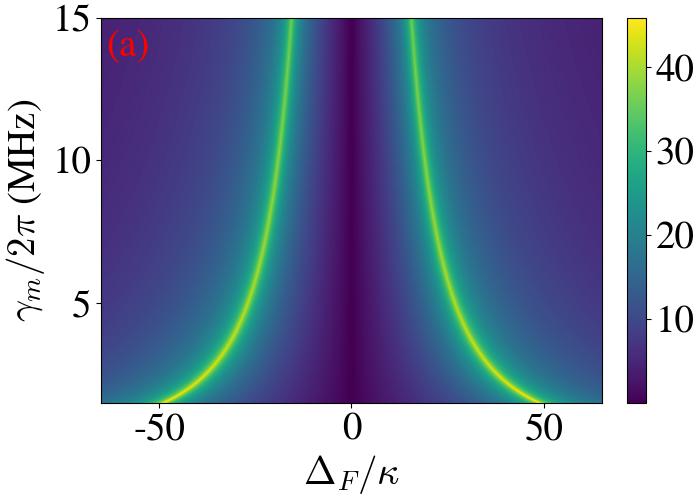}
        \label{fig:2a}
    \end{minipage}\hfill

    \begin{minipage}[t]{0.88\columnwidth}
        \centering
        \includegraphics[width=\textwidth]{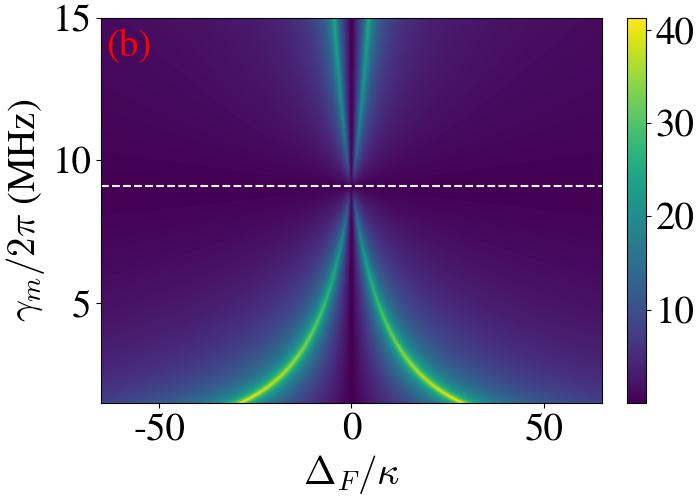}

        \label{fig:2b}
    \end{minipage}
    \caption{The isolation ratio $I$ vs $\Delta_F/\kappa$ and $\gamma_m$. $\Delta/\kappa=0$ in panel (a) and 20 in panel (b). $\kappa /2\pi$ = 1.1 MHz. The other parameters are the same as in Fig. \ref{fig:fig2}. The plot is symmetric with respect to the line $\Delta_F/\kappa=0$. The maximum values of $I$ are 45.890 dB (when $\gamma_m=1.5$ MHz) in panel (a) and 41.352 dB (when $\gamma_m=1.503$MHz) in panel (b).}
    \label{fig:fig3}
\end{figure}

As seen in Fig. \ref{fig:fig2}(a), at $\Delta_F/\gamma_m=8.295$ and $\Delta_F/\gamma_m=-8.295$, the ratio of the two transmission coefficients is maximized, and consequently, the nonreciprocal transmission effect is most pronounced. Correspondingly, in Fig. \ref{fig:fig2}(b), the isolation ratio $I$ reaches an identical maximum value of 41.63 dB at both these points. The inset in Fig. \ref{fig:fig2}(a) shows that the forward transmission coefficient reaches a high of 0.67, with the backward transmission coefficient being only 0.01. Importantly, this pronounced asymmetry in transmission is maintained stably within a specific range of $\Delta_F/\gamma_m$. A similar stability is also observed for the isolation ratio, as shown in the inset of Fig. \ref{fig:fig2}(b). Additionally, for $\Delta_F/\gamma_m =0$, we find $T_{21}$ = $T_{12}$. This condition occurs when the WGM cavity is not rotating, causing $\Delta_1 = \Delta_2 $ and thus eliminating nonreciprocity.
\begin{figure}[htbp]
	\centering
    \vspace{0.1em}
	\begin{minipage}[t]{0.88\columnwidth}
		\centering
		\includegraphics[width=\textwidth]{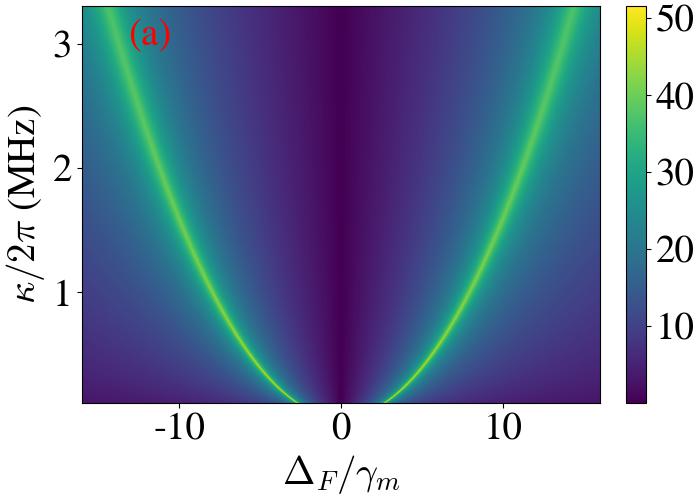}

	\end{minipage}%
	\hfill

	\begin{minipage}[t]{0.88\columnwidth}
		\centering
		\includegraphics[width=\textwidth]{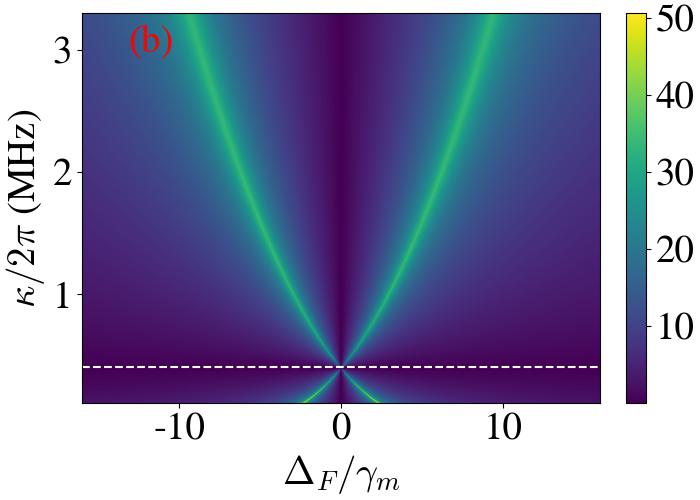}

	\end{minipage}

    \caption{The isolation ratio $I$ vs $\Delta_F/\gamma_m$ and $\kappa$. $\Delta/\gamma_m=0$ in panel (a) and 5 in panel (b). $\gamma_m/2\pi=4$ MHz. The other parameters are the same as in Fig. \ref{fig:fig2}. The maximum values of $I$ are 51.437 dB (when $\kappa=0.114$MHz) in panel (a) and 50.542 dB (when $\kappa=0.112$MHz) in panel (b).}
    \label{fig:fig4}
\end{figure}

We now turn our attention to investigating the impact of $\gamma_m$ on the isolation ratio $I$.  As can be seen from Fig. \ref{fig:fig3}(a), for any given value of $\Delta_F$, an optimal value of $\gamma_m$ can always be found that maximizes the isolation ratio $I$. This maximum achievable $I$ can be sustained at a high level of around 40 dB. In particular, $I = 0$ when $\Delta_F/\kappa = 0$, which corresponds to reciprocal transmission. Furthermore, the entire plot is symmetric about the axis $\Delta_F/\kappa = 0$. The region where $\Delta_F/\kappa>0$ corresponds to the case $|A_{1,\mathrm{out}}|^2 > |A_{2,\mathrm{out}}|^2$, while the region where $\Delta_F/\kappa<0$ corresponds to $|A_{1,\mathrm{out}}|^2 < |A_{2,\mathrm{out}}|^2$; this characteristic is also present in Fig. \ref{fig:fig2}(a). If we reverse the rotational direction of the WGM cavity, the physical meaning of the regions on either side of the symmetry axis is also interchanged (i.e., the region where $\Delta_F/\kappa>0$ then corresponds to $|A_{1,\mathrm{out}}|^2 < |A_{2,\mathrm{out}}|^2$ and $\Delta_F/\kappa<0$ to $|A_{1,\mathrm{out}}|^2 > |A_{2,\mathrm{out}}|^2$). Therefore, we can reverse the direction of nonreciprocal transmission simply by reversing the WGM cavity's rotation.

\begin{figure}[ht]
	\centering
    \vspace{0.4em}
	\begin{minipage}[t]{0.88\columnwidth}
		\centering
		\includegraphics[width=\textwidth]{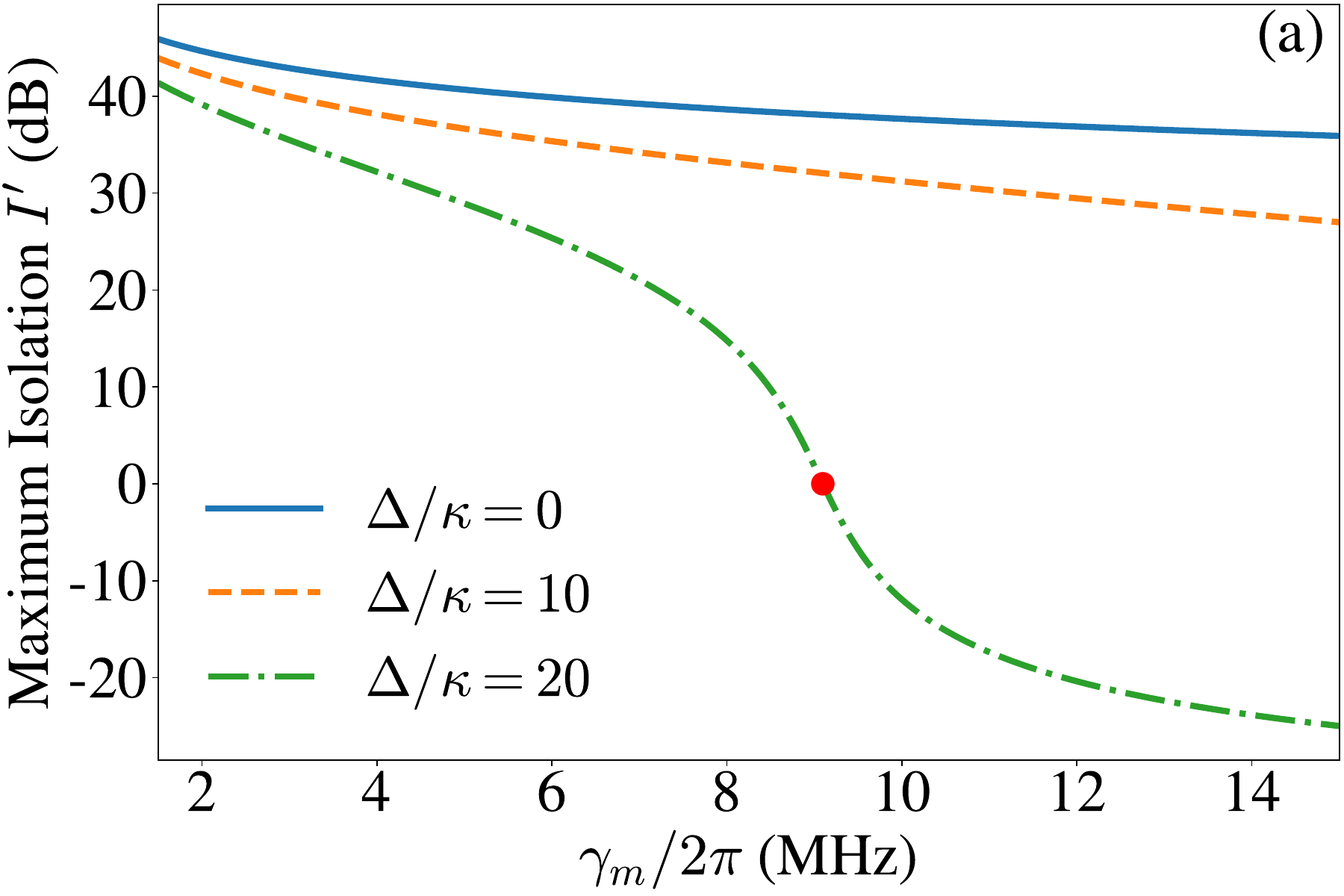}

	\end{minipage}%
	\hfill

	\begin{minipage}[t]{0.88\columnwidth}
		\centering
		\includegraphics[width=\textwidth]{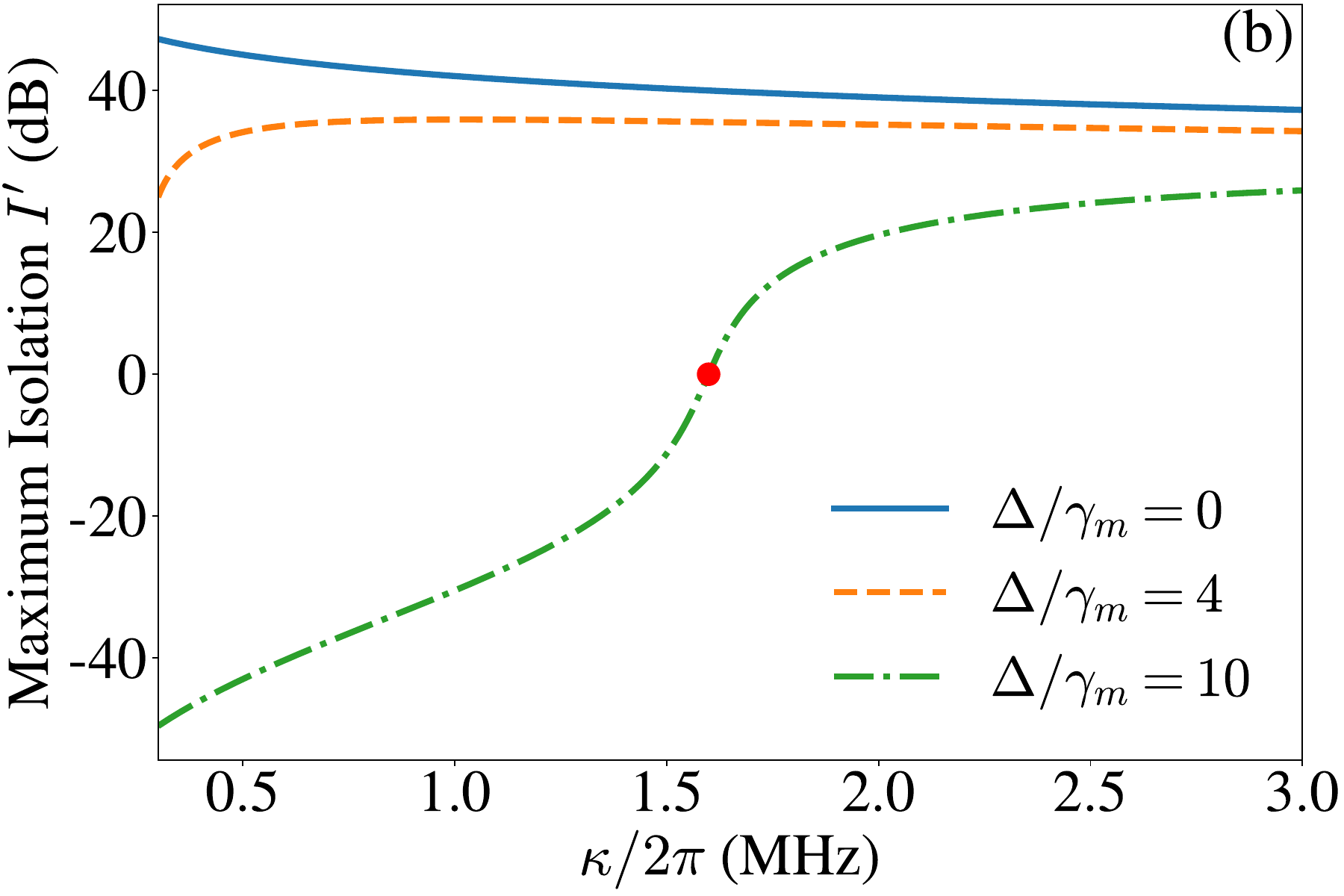}

	\end{minipage}

    \caption{Under optimal parameter $\Delta_F$ conditions, the isolation ratio $I$ varies with (a) $\gamma_m$ and (b) $\kappa$. $\kappa/2\pi=1.1$ MHz in panel (a) and $\gamma_m/2\pi=4$ MHz in panel (b). The other parameters are the same as in Fig. \ref{fig:fig2}. The red dots represent the reciprocal points $\gamma_0$ and $\kappa_0$. Indeed, this functional plot corresponds to Eq. (\ref{eq18}).}
    \label{fig:fig5}
\end{figure}

When $\Delta/\kappa \ne 0$, as shown in Fig. \ref{fig:fig3}(b), we surprisingly find that when the system satisfies the condition $\Delta=g\sqrt{\frac{\kappa}{\gamma_m}}$, termed the impedance matching condition \cite{PhysRevApplied.12.034001,PhysRevA.98.063845}, the isolation ratio $I$ becomes zero, corresponding to the white dashed line in the figure. We define the $\gamma_m$ value satisfying the impedance matching condition as the reciprocal point $\gamma_0$. As observed from Eqs. (\ref{eq13}) and (\ref{eq14}), the satisfaction of the reciprocity condition by the system parameters leads to $|{A_{1,\mathrm{out}}}|^2={|A_{2,\mathrm{out}}|}^2$, under which $I$ remains identically zero. Furthermore, in contrast to the $\Delta/\kappa=0$ situation, when $\gamma_m>\gamma_0$, the system exhibits a reversal in the direction of nonreciprocal transmission. Specifically, in the regions where $\Delta_F/\kappa>0$ ($\Delta_F/\kappa<0$) and $\gamma_m>\gamma_0$, we observe $|A_{1,\mathrm{out}}|^2 < |A_{2,\mathrm{out}}|^2$ ($|A_{1,\mathrm{out}}|^2 > |A_{2,\mathrm{out}}|^2$). Naturally, we can expect the system's response to $\kappa$ to mirror its response to $\gamma_m$ in many respects, as illustrated in Fig. \ref{fig:fig4}. We can also define a $\kappa$-dependent reciprocal point $\kappa_0$ via the impedance-matching condition. In contrast to $\gamma_m$, a reversal of the nonreciprocal transmission direction occurs when $\kappa<\kappa_0$. Consequently, by analyzing Figs. \ref{fig:fig3} and \ref{fig:fig4}, we can find the $\kappa$ and $\gamma_m$ values that yield the maximum isolation $I$ for a given $g$. Importantly, this maximum $I$ can remain at a large and stable value over a certain range of parameter variations.

\begin{figure}[ht]
	\centering

	\begin{minipage}[t]{0.88\columnwidth}
		\centering
		\includegraphics[width=\textwidth]{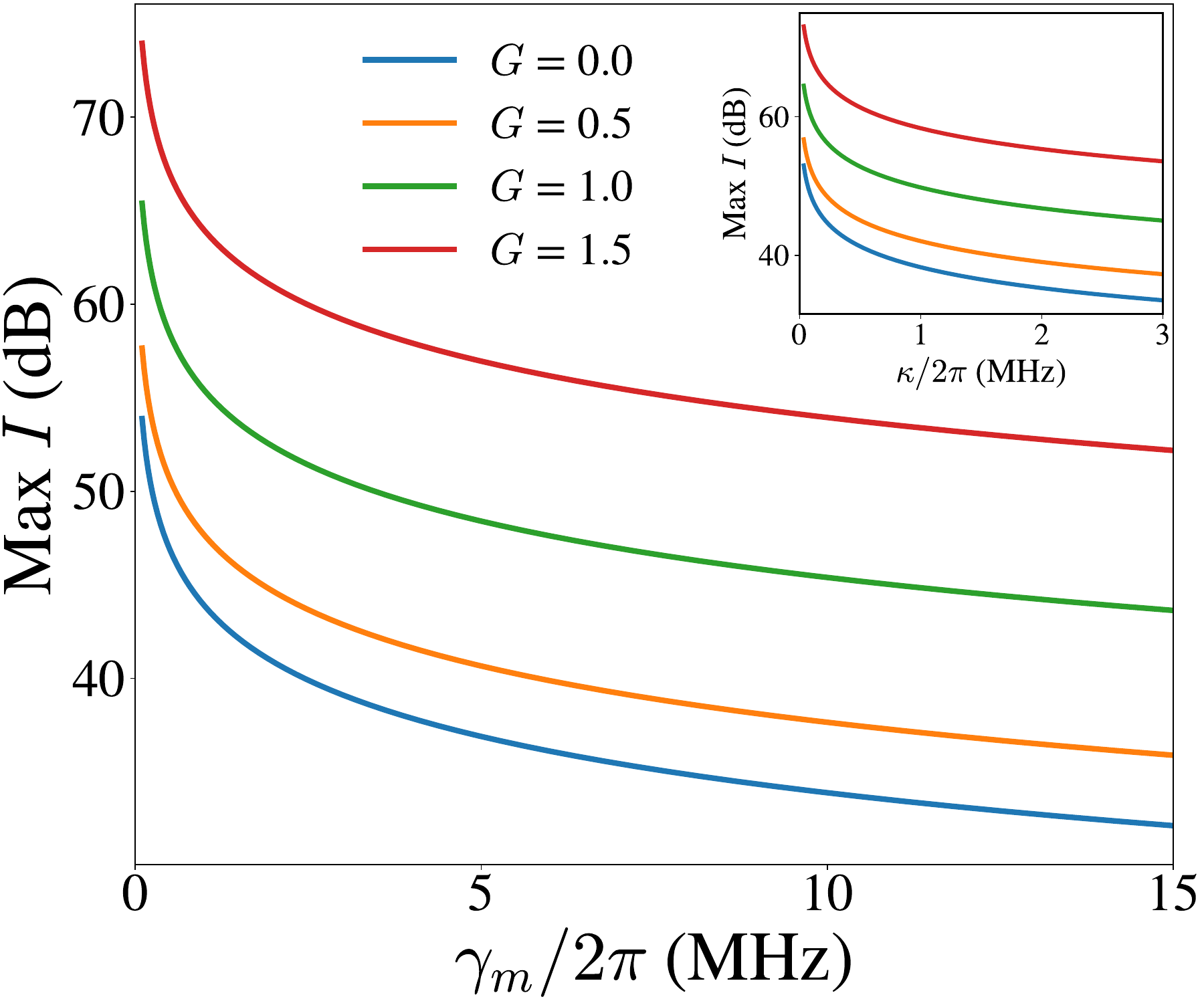}

	\end{minipage}%

    \caption{The maximum isolation ratio $I$ as a function of $\gamma_m$ (main panel) and $\kappa$ (inset) for different $G$ values, under optimal $\Delta_F$ selection. $\kappa/2\pi=1.1$ MHz in the main plot and $\gamma_m/2\pi= 4$ MHz in the inset. The other parameters are the same as in Fig. \ref{fig:fig2}. The line colors used in the inset are consistent with the different $G$ values presented in the main plot.}
    \label{fig:fig6}
\end{figure}

Next, we further investigate the relationship between the isolation ratio and $\gamma_m$, as well as $\kappa$. Please note that to clearly illustrate the change in nonreciprocal transmission direction, the vertical axis in Fig. \ref{fig:fig5} does not adopt the definition from Eq. (\ref{eq16}). Instead, we define $I'=10 \log_{10} \frac{|A_{1,\mathrm{out}}|^2}{|A_{2,\mathrm{out}}|^2}$. Consequently, when $I'<0$, it signifies a reversal of the nonreciprocal transmission direction. As shown in Fig. \ref{fig:fig5}, under the condition of optimal parameter $\Delta_F$, the maximum achievable value of $I'$ increases monotonically as either $\gamma_m$ or $\kappa$ decreases when $\Delta=0$. For $\gamma_m$, this can be readily understood: A larger $\gamma_m$ leads to stronger dissipation of the magnon mode, hindering efficient signal transmission and thus reducing $I'$. As for $\kappa$, a larger value enhances the escape efficiency of photons, which would increase the transmission coefficients. However, the forward transmission coefficient has already reached its maximum and remains almost unaffected by further increases in $\kappa$, resulting in a decrease in $I'$. Despite the increase in dissipation, we observe that $I'$ exhibits minimal fluctuations and is stably maintained at a level exceeding 40 dB. This highlights the system's significant robustness to dissipation.

\begin{figure}[ht]
	\centering
    \hfill
	\begin{minipage}[t]{0.88\columnwidth}
		\centering
		\includegraphics[width=\textwidth]{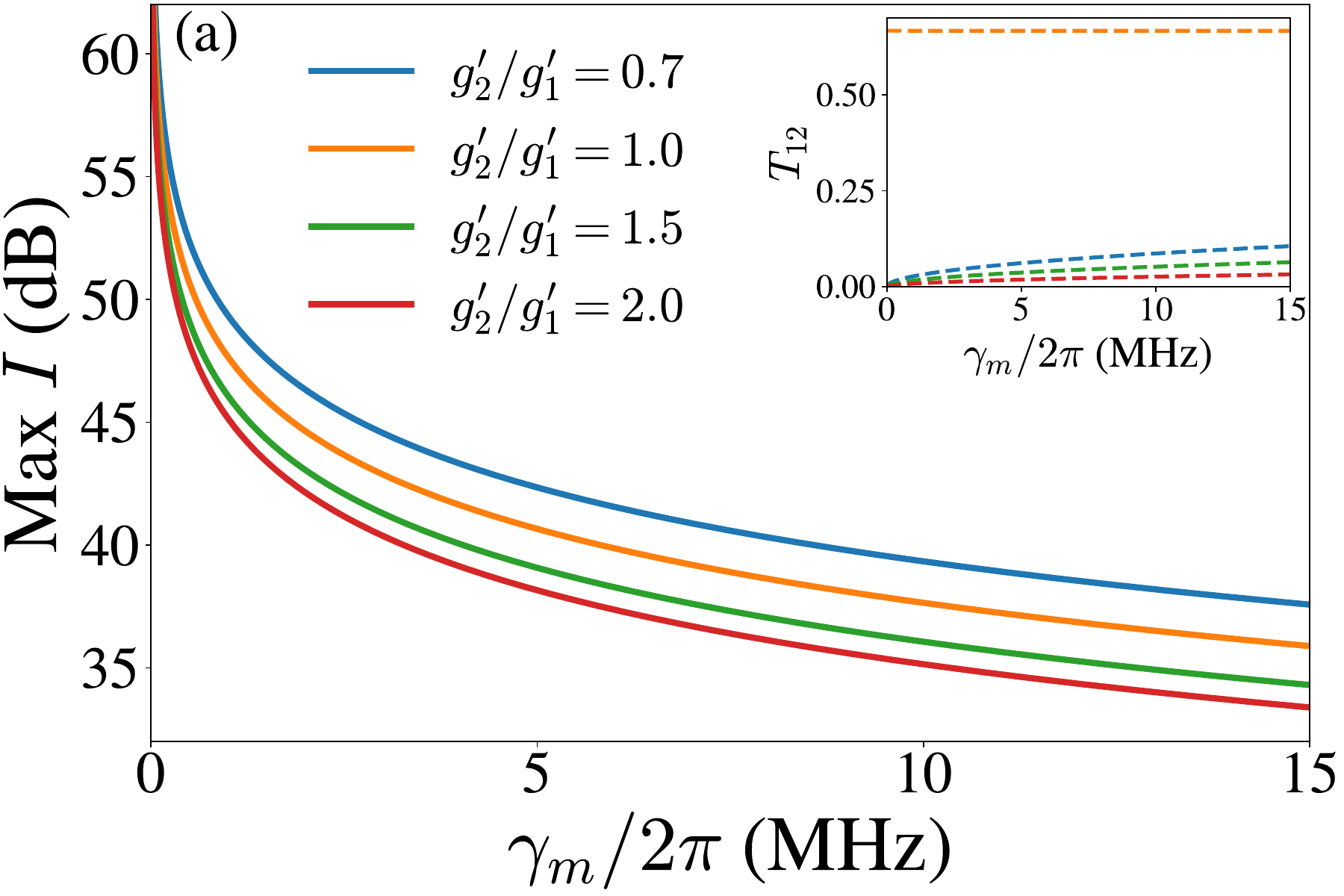}
	\end{minipage}%
	\hfill
    \vspace{0.1em}
    \begin{minipage}[t]{0.88\columnwidth}
		\centering
		\includegraphics[width=\textwidth]{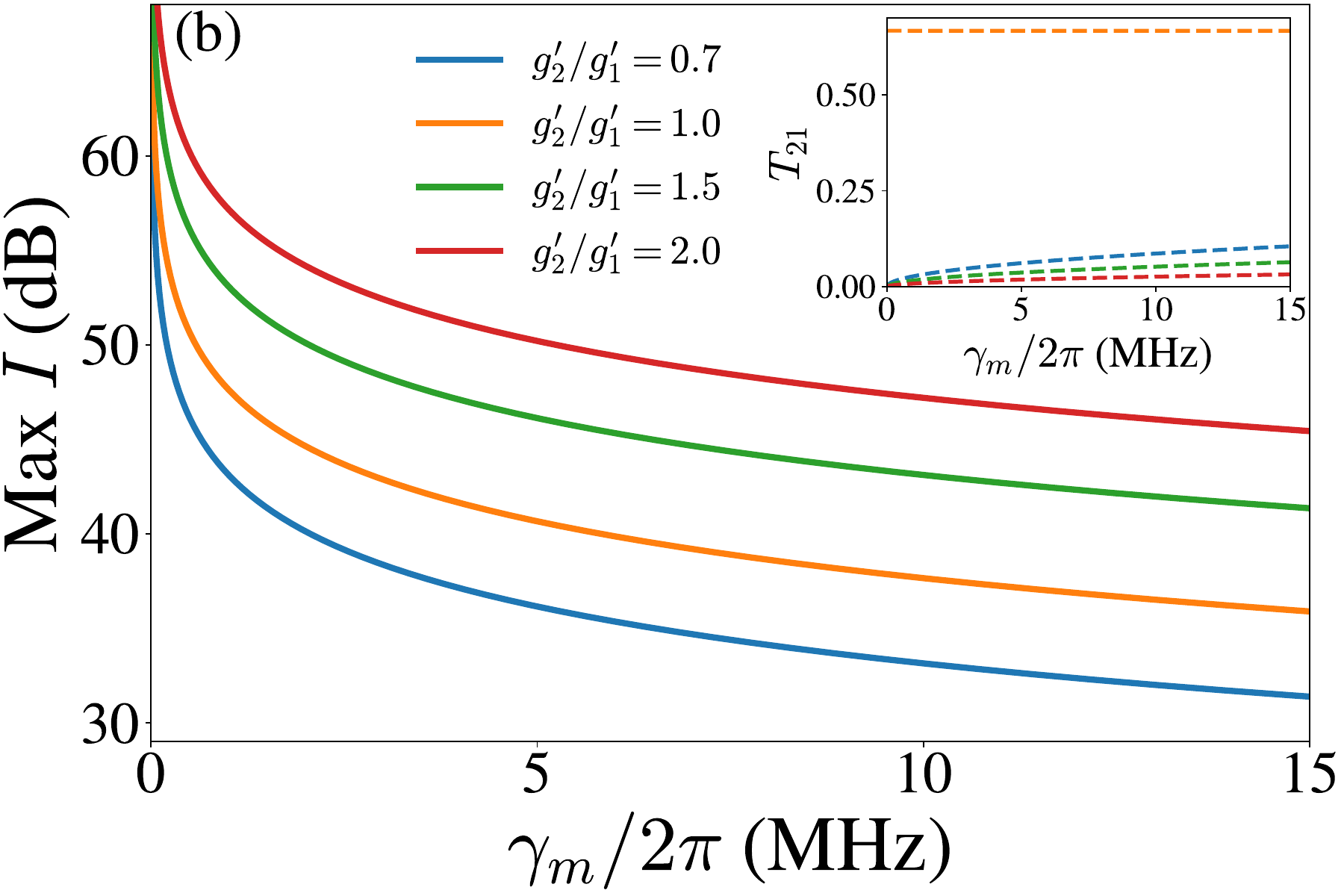}
	\end{minipage}%

    \caption{The maximum achievable value of $I$ at different $g_2^{'}/g_1^{'}$ values varies with $\gamma_m$ for (a) $\Delta_F>0$ and (b) $\Delta_F<0$. $\kappa/2\pi=1.1$ MHz, $G=0.5$, and $g_1/2\pi=41$ MHz. The other parameters are the same as in Fig. \ref{fig:fig2}. The insets in panel (a) and (b), respectively, show the variation of $T_{12}$ and $T_{21}$ with $\gamma_m$ at the points where $I$ reaches its maximum. The line colors used in the insets are consistent with the different $g_2^{'}/g_1^{'}$ values presented in the main plot.}
    \label{fig:fig7}
\end{figure}

When $\Delta \ne 0$, in the regime where $\gamma_m<\gamma_0$, the maximum achievable $I'$ exhibits a monotonic increase with decreasing $\gamma_m$, as depicted in Fig. \ref{fig:fig5}(a). Nevertheless, when $\gamma_m>\gamma_0$, the system's nonreciprocal transmission direction reverses (i.e., $I'<0$). Moreover, the absolute value of $I'$ increases as $\gamma_m$ increases. While this seems to contradict our earlier discussion, an examination of the system's transmission coefficients reveals a significant drop to very low values. For instance, at $\Delta/\kappa=20$ and $\gamma_m/2\pi=12$ MHz, we obtain $T_{21}=0.0038$ and $T_{12}=0.0004$. The extremely low transmission coefficients in this regime render information transmission nearly impossible. Furthermore, the forward transmission isolation decays rapidly as $\gamma_m$ increases. Consider, for example, the cases where $\Delta/\kappa=10$. When $\gamma_m/2\pi=2$ MHz, $T_{21}=0.0003$ and $T_{12}=0.0382$. When $\gamma_m/2\pi=4$ MHz, $T_{21}=0.0002$ and $T_{12}=0.0084$. The reason is that a larger $\Delta$ leads to a ``dilute'' of the nonreciprocity induced by the Sagnac effect, consequently weakening the nonreciprocal effect. On the other hand, detuning $\Delta$ makes it more difficult for the signal to couple into the optical cavity. Moreover, a larger $\gamma_m$ prevents photons from the input port from effectively transferring to the output port via coupling with magnons. Both of these factors lead to a rapid decrease in transmission coefficients. This also compromises the system's robustness, making it more sensitive to dissipation. When $\gamma_m$ increases, the forward transmission coefficient $T_{12}$ exhibits greater sensitivity to $\gamma_m$ compared to the reverse transmission coefficient $T_{21}$. Consequently, $T_{12}$ decreases more rapidly until the impedance matching condition is satisfied. If $\gamma_m$ is further increased, a reversal of the nonreciprocal transmission will occur. Therefore, it is necessary to keep $\Delta=0$ for practical parameter settings. The same analysis is applicable to $\kappa$. The sole distinction is that the regime of reversed nonreciprocal transmission and low transmission coefficients occurs when $\kappa<\kappa_0$, as illustrated in Fig. \ref{fig:fig5}(b). In contrast to $\gamma_m$, a reduction in $\kappa$ leads to a decrease in transmission coefficients. This is attributed to the fact that a smaller $\kappa$ signifies a lower probability for photons to exit the cavity.

We now focus on investigating the impact of $G$ on the nonreciprocal transmission efficiency of the system. Figure \ref{fig:fig6} shows the dependence of the maximum achievable isolation $I$ on $\gamma_m$ and $\kappa$ for the optimal $\Delta_F$. As illustrated in the main plot, there is a direct relationship where larger $G$ values correspond to higher achievable $I$ values, which lead to a significant enhancement in nonreciprocal transmission efficiency. For instance, with $G = 1$ and $\gamma_m/2\pi=4$ MHz, $I$ can exceed 50 dB, signifying almost perfect nonreciprocal transmission. Crucially, even with $G=0$ (i.e., no squeeze applied), a substantial isolation ratio can still be obtained. This also shows that the system's nonreciprocity is independent of whether the magnon mode is squeezed. Similarly, the same conclusion holds for the analysis of $\kappa$, as depicted in the inset.

Since the previous discussion focused on a relatively specific situation $g_1^{'}=g_2^{'}$, it is essential to address a more general scenario here, such as $g_1^{'} \neq g_2^{'}$. At this point, Eq. (\ref{eq17}) is no longer valid. The extremum points of $R$ satisfy the following equation:
\begin{align}
&\Delta_{F_{1,2}} = \frac{1}{2} \left[ U_1 \pm \sqrt{ {U_1}^2 + U_2 } \right], \label{eq19}\nonumber \\
&U_1 = \sqrt{\frac{\kappa}{\gamma_m}}(g_1' - g_2') \nonumber \\
&U_2 = \kappa^2 + 4\left(\Delta - g_1'\sqrt{\frac{\kappa}{\gamma_m}}\right)\left(\Delta - g_2'\sqrt{\frac{\kappa}{\gamma_m}}\right)\nonumber \\
&R(\Delta_{F_{1,2}}) = - \left( \frac{g_1'}{g_2'} \right)^2 \frac{\Delta - \Delta_{F_{1,2}} - g_2'\sqrt{\frac{\kappa}{\gamma_m}}}{\Delta + \Delta_{F_{1,2}} - g_1'\sqrt{\frac{\kappa}{\gamma_m}}}.
\end{align}
Under the condition $g_1^{'} \neq g_2^{'}$, it is usually the case that $\Delta_{F_{1}} \neq \Delta_{F_{2}}$. Consequently, the nonreciprocal transmission property, exhibiting symmetry around $\Delta_{F}=0$, is broken. Perfectly symmetric reverse transmission efficiency cannot be achieved by simply reversing the WGM cavity's rotation. As depicted in Fig. \ref{fig:fig7}(a), for $\Delta_F>0$ (corresponding to $|A_{1,\mathrm{out}}|^2 > |A_{2,\mathrm{out}}|^2$), the maximum achievable $I$ is plotted as a function of $\gamma_m$ for various $g_2^{'}/g_1^{'}$ ratios. It can be found that a smaller ratio of $g_2^{'}/g_1^{'}$ leads to a higher forward transmission isolation. This observation is readily understood by analyzing the expression
\begin{align}
R=\left|\frac{-ig_1^{'}D_2F_3-g_1^{'}g_2^{'}F_2}{-ig_2^{'}D_1F_3-g_1^{'}g_2^{'}F_1}\right|^2.
\end{align}
With $g_1^{'}$ held constant, a reduction in $g_2^{'}$ results in a larger $R$ value, which consequently enhances the forward transmission isolation of the system. When $\Delta_F<0$ (corresponding to $|A_{1,\mathrm{out}}|^2 < |A_{2,\mathrm{out}}|^2$), the situation is reversed, whereby an increase in $g_2^{'}$ enhances the system's reverse transmission isolation, a behavior demonstrated in Fig. \ref{fig:fig7}(b). However, calculating the transmission coefficients reveals significant attenuation, as shown in the insets of Figs. \ref{fig:fig7}(a) and \ref{fig:fig7}(b). Moreover, this attenuation becomes more severe as the difference between $g_1^{'}$ and $g_2^{'}$ increases. For example, at $\Delta_F<0$ and $\gamma_m/2\pi= 4$ MHz, $T_{21}$ is 0.033 when $g_2^{'}/g_1^{'}=1.5$, and 0.017 when $g_2^{'}/g_1^{'}=2.0$. Therefore, it is necessary to maintain $g_1^{'} = g_2^{'}$ in practical parameter tuning. The analysis for $\gamma_m$ is directly transferable to $\kappa$. For the sake of brevity, further discussion is omitted. Drawing from the preceding discussions, we see that with well-chosen system parameters, an isolation ratio exceeding 40 dB is easily attainable. This is a significant step toward realizing high-isolation nonreciprocal optical transmission systems.

\section{DISCUSSION AND CONCLUSIONS} \label{IV}
Let us give a brief discussion of the feasibility of the experiment. The magnon frequency $\omega_m$ is experimentally controlled by tuning the external bias magnetic field $H$ \cite{doi:10.1126/sciadv.1501286}. For a cavity-magnon system, a typical parameter choice is a cavity resonant frequency and magnon frequency of $\omega_0 =193$ THz and $\omega_m = 10.1$ GHz \cite{Ren:21,Kong:21,PhysRevLett.120.133602,PhysRevLett.117.123605,PhysRevApplied.12.034001,PhysRevLett.114.080503}. The value $g_0 = 41$ MHz represents a typical cavity-magnon coupling strength employed in both experimental and numerical studies \cite{zhang2019theory,PhysRevLett.120.057202,PhysRevResearch.1.023021,PhysRevLett.134.196904} and the squeezing parameter $G$ typically ranges from 0 to 3 \cite{PhysRevA.100.062501,PhysRevResearch.3.023126,PhysRevA.111.053707,Wang_2025}. $\kappa/2 \pi \sim 0.1-100$ MHz and $\gamma_m$ is of the same order of magnitude \cite{RevModPhys.86.1391,doi:10.1126/sciadv.1501286,Mathkoor2025,PhysRevLett.121.087205,PhysRevA.109.043512}. For a rotating WGM cavity, typical parameter values can be taken as $n=2.2$, $r=1.1$ mm, and $\Omega/2\pi=6.6$ kHz \cite{Maayani2018,10.1063/5.0190162}. For this parameter set, Eq. (\ref{eq1}) yields $\Delta_F \approx 65$ MHz, where we have considered only the first term because the other two are typically negligible. The magnitude of $\Delta_F$ is crucially limited by the fact that mechanical rotation angular velocity is typically below 10 kHz. Consequently, overcoming this limitation in the future would lead to a broader tunable parameter range.

In summary, we have theoretically proposed a scheme utilizing a cavity-magnon system to achieve ultrahigh nonreciprocal optical transmission through the exploitation of the Sagnac effect. In a rotating WGM cavity, the Fizeau shift induced by the Sagnac effect leads to nonreciprocal optical transmission. Our analysis demonstrates that, under experimentally feasible parameters, it is possible to attain nonreciprocal transmission with a high isolation ratio exceeding 40 dB, which also exhibits significant robustness against variations in angular velocity $\Omega$ and dissipation. Moreover, the optical isolation ratio can be further enhanced by increasing the squeezing parameter $G$. A notable feature of this system is the reconfigurability of the transmission direction, which can be readily reversed by simply changing the rotation direction of the resonator, without requiring any modifications to other system parameters. The proposed nonreciprocal optical transmission system offers several advantages, including a high isolation ratio, compact form factor, and operational simplicity. These attributes collectively make it a promising candidate for realizing on-chip optical isolators with enhanced signal-to-noise performance.

\bibliography{citation}

@misc{barzanjeh2025nonreciprocityquantumtechnology,
      title={Nonreciprocity in Quantum Technology}, 
      author={Shabir Barzanjeh and André Xuereb and Andrea Alù and Sander A. Mann and Nikita Nefedkin and Vittorio Peano and Peter Rabl},
      year={2025},
      eprint={2508.03945},
      archivePrefix={arXiv},
      primaryClass={quant-ph},
      url={https://arxiv.org/abs/2508.03945}, 
}

@article{sounas2017non,
  title={Non-reciprocal photonics based on time modulation},
  author={Sounas, Dimitrios L and Alù, Andrea},
  journal={Nat. Photonics},
  volume={11},
  number={12},
  pages={774--783},
  year={2017},
  publisher={Nature Publishing Group UK London},
  url={https://doi.org/10.1038/s41566-017-0051-x},
  doi={10.1038/s41566-017-0051-x},
}

@article{Shui:22,
author = {Tao Shui and Wen-Xing Yang and Mu-Tian Cheng and Ray-Kuang Lee},
journal = {Opt. Express},
number = {4},
pages = {6284--6299},
publisher = {Optica Publishing Group},
title = {Optical nonreciprocity and nonreciprocal photonic devices with directional four-wave mixing effect},
volume = {30},
year = {2022},
url = {https://opg.optica.org/oe/abstract.cfm?URI=oe-30-4-6284},
doi = {10.1364/OE.446238},
}

@article{PhysRevA.106.053707,
  title = {Nonreciprocal photon blockade in cavity optomagnonics},
  author = {Xie, Hong and He, Le-Wei and Shang, Xiao and Lin, Gong-Wei and Lin, Xiu-Min},
  journal = {Phys. Rev. A},
  volume = {106},
  issue = {5},
  pages = {053707},
  numpages = {9},
  year = {2022},
  month = {Nov},
  publisher = {American Physical Society},
  doi = {10.1103/PhysRevA.106.053707},
  url = {https://link.aps.org/doi/10.1103/PhysRevA.106.053707}
}

@book{mariz2022lorentz,
  title={Lorentz symmetry breaking--classical and quantum aspects},
  author={Tiago Mariz, Jose Roberto Nascimento, Albert Petrov},
  year={2023},
  series={SpringerBriefs in Physics},
  publisher={Springer},
  doi = {10.1007/978-3-031-20120-2},
  url = {https://doi.org/10.1007/978-3-031-20120-2},
archivePrefix = {arXiv},
  eprint = {2205.02594},
  primaryClass = {hep-th}
}

@article{jalas2013and,
  title={What is—and what is not—an optical isolator},
  author={Jalas, Dirk and Petrov, Alexander and Eich, Manfred and Freude, Wolfgang and Fan, Shanhui and Yu, Zongfu and Baets, Roel and Popovi{\'c}, Milo{\v{s}} and Melloni, Andrea and Joannopoulos, John D and others},
  journal={Nat. Photonics},
  volume={7},
  number={8},
  pages={579--582},
  year={2013},
  publisher={Nature Publishing Group UK London},
doi = {https://doi.org/10.1038/nphoton.2013.185},
  url = {10.1038/nphoton.2013.185}
}

@article{
doi:10.1126/science.aam6662,
author = {K. L. Tsakmakidis  and L. Shen  and S. A. Schulz  and X. Zheng  and J. Upham  and X. Deng  and H. Altug  and A. F. Vakakis  and R. W. Boyd },
title = {Breaking Lorentz reciprocity to overcome the time-bandwidth limit in physics and engineering},
journal = {Science},
volume = {356},
number = {6344},
pages = {1260-1264},
year = {2017},
doi = {10.1126/science.aam6662},
URL = {https://www.science.org/doi/abs/10.1126/science.aam6662},
}

@article{10.1063/5.0166231,
    author = {Wang, Zi-Yuan and Qian, Jie and Wang, Yi-Pu and Li, Jie and You, J. Q.},
    title = {Realization of the unidirectional amplification in a cavity magnonic system},
    journal = {Appl. Phys. Lett.},
    volume = {123},
    number = {15},
    pages = {153904},
    year = {2023},
    issn = {0003-6951},
    doi = {10.1063/5.0166231},
    url = {https://doi.org/10.1063/5.0166231},
}

@article{PhysRevLett.119.147703,
  title = {Breaking Lorentz Reciprocity with Frequency Conversion and Delay},
  author = {Rosenthal, Eric I. and Chapman, Benjamin J. and Higginbotham, Andrew P. and Kerckhoff, Joseph and Lehnert, K. W.},
  journal = {Phys. Rev. Lett.},
  volume = {119},
  issue = {14},
  pages = {147703},
  numpages = {5},
  year = {2017},
  month = {Oct},
  publisher = {American Physical Society},
  doi = {10.1103/PhysRevLett.119.147703},
  url = {https://link.aps.org/doi/10.1103/PhysRevLett.119.147703}
}

@article{Aplet:64,
author = {L. J. Aplet and J. W. Carson},
journal = {Appl. Opt.},
number = {4},
pages = {544--545},
publisher = {Optica Publishing Group},
title = {A Faraday Effect Optical Isolator},
volume = {3},
month = {Apr},
year = {1964},
url = {https://opg.optica.org/ao/abstract.cfm?URI=ao-3-4-544},
doi = {10.1364/AO.3.000544},
}

@article{PhysRevA.92.063845,
  title = {Magneto-optical coupling in whispering-gallery-mode resonators},
  author = {Haigh, J. A. and Langenfeld, S. and Lambert, N. J. and Baumberg, J. J. and Ramsay, A. J. and Nunnenkamp, A. and Ferguson, A. J.},
  journal = {Phys. Rev. A},
  volume = {92},
  issue = {6},
  pages = {063845},
  numpages = {7},
  year = {2015},
  month = {Dec},
  publisher = {American Physical Society},
  doi = {10.1103/PhysRevA.92.063845},
  url = {https://link.aps.org/doi/10.1103/PhysRevA.92.063845}
}

@article{https://doi.org/10.1002/lpor.201900252,
author = {Chai, Cheng-Zhe and Zhao, Hao-Qi and Tang, Hong X. and Guo, Guang-Can and Zou, Chang-Ling and Dong, Chun-Hua},
title = {Non-Reciprocity in High-Q Ferromagnetic Microspheres via Photonic Spin–Orbit Coupling},
journal = {Laser Photonics Rev.},
volume = {14},
number = {2},
pages = {1900252},
doi = {https://doi.org/10.1002/lpor.201900252},
url = {https://onlinelibrary.wiley.com/doi/abs/10.1002/lpor.201900252},
year = {2020}
}

@article{Nazari:14,
author = {F. Nazari and N. Bender and H. Ramezani and M.K. Moravvej-Farshi and D. N. Christodoulides and T. Kottos},
journal = {Opt. Express},
number = {8},
pages = {9574--9584},
publisher = {Optica Publishing Group},
title = {Optical isolation via PT-symmetric nonlinear Fano resonances},
volume = {22},
month = {Apr},
year = {2014},
url = {https://opg.optica.org/oe/abstract.cfm?URI=oe-22-8-9574},
doi = {10.1364/OE.22.009574},
}

@article{PhysRevLett.121.123601,
  title = {Nonreciprocity Realized with Quantum Nonlinearity},
  author = {Rosario Hamann, Andr\'es and M\"uller, Clemens and Jerger, Markus and Zanner, Maximilian and Combes, Joshua and Pletyukhov, Mikhail and Weides, Martin and Stace, Thomas M. and Fedorov, Arkady},
  journal = {Phys. Rev. Lett.},
  volume = {121},
  issue = {12},
  pages = {123601},
  numpages = {5},
  year = {2018},
  month = {Sep},
  publisher = {American Physical Society},
  doi = {10.1103/PhysRevLett.121.123601},
  url = {https://link.aps.org/doi/10.1103/PhysRevLett.121.123601}
}

@article{
doi:10.1126/science.1214383,
author = {Li Fan  and Jian Wang  and Leo T. Varghese  and Hao Shen  and Ben Niu  and Yi Xuan  and Andrew M. Weiner  and Minghao Qi },
title = {An All-Silicon Passive Optical Diode},
journal = {Science},
volume = {335},
number = {6067},
pages = {447-450},
year = {2012},
doi = {10.1126/science.1214383},
URL = {https://www.science.org/doi/abs/10.1126/science.1214383},
}

@article{yu2009complete,
  title={Complete optical isolation created by indirect interband photonic transitions},
  author={Yu, Zongfu and Fan, Shanhui},
  journal={Nat. Photonics},
  volume={3},
  number={2},
  pages={91--94},
  year={2009},
  publisher={Nature Publishing Group UK London},
  doi = {10.1038/nphoton.2008.273},
  url = {https://doi.org/10.1038/nphoton.2008.273}
}

@article{PhysRevLett.110.093901,
  title = {Optical Diode Made from a Moving Photonic Crystal},
  author = {Wang, Da-Wei and Zhou, Hai-Tao and Guo, Miao-Jun and Zhang, Jun-Xiang and Evers, J\"org and Zhu, Shi-Yao},
  journal = {Phys. Rev. Lett.},
  volume = {110},
  issue = {9},
  pages = {093901},
  numpages = {5},
  year = {2013},
  month = {Feb},
  publisher = {American Physical Society},
  doi = {10.1103/PhysRevLett.110.093901},
  url = {https://link.aps.org/doi/10.1103/PhysRevLett.110.093901}
}

@article{
doi:10.1126/science.1246957,
author = {Romain Fleury  and Dimitrios L. Sounas  and Caleb F. Sieck  and Michael R. Haberman  and Andrea Alù },
title = {Sound Isolation and Giant Linear Nonreciprocity in a Compact Acoustic Circulator},
journal = {Science},
volume = {343},
number = {6170},
pages = {516-519},
year = {2014},
doi = {10.1126/science.1246957},
URL = {https://www.science.org/doi/abs/10.1126/science.1246957},
}

@article{PhysRevA.91.053854,
  title = {Optical nonreciprocity and optomechanical circulator in three-mode optomechanical systems},
  author = {Xu, Xun-Wei and Li, Yong},
  journal = {Phys. Rev. A},
  volume = {91},
  issue = {5},
  pages = {053854},
  numpages = {8},
  year = {2015},
  month = {May},
  publisher = {American Physical Society},
  doi = {10.1103/PhysRevA.91.053854},
  url = {https://link.aps.org/doi/10.1103/PhysRevA.91.053854}
}

@article{yan2019perfect,
  title={Perfect optical nonreciprocity in a double-cavity optomechanical system},
  author={Yan, Xiao-Bo and Lu, He-Lin and Gao, Feng and Yang, Liu},
  journal={Front. Phys.},
  volume={14},
  pages={1--6},
  year={2019},
  publisher={Springer},
  doi = {10.1007/s11467-019-0922-3},
  url = {https://doi.org/10.1007/s11467-019-0922-3}
}

@article{PhysRevA.98.063845,
  title = {Optomechanically induced nonreciprocity in a three-mode optomechanical system},
  author = {Xu, Xun-Wei and Song, L. N. and Zheng, Qiang and Wang, Z. H. and Li, Yong},
  journal = {Phys. Rev. A},
  volume = {98},
  issue = {6},
  pages = {063845},
  numpages = {7},
  year = {2018},
  month = {Dec},
  publisher = {American Physical Society},
  doi = {10.1103/PhysRevA.98.063845},
  url = {https://link.aps.org/doi/10.1103/PhysRevA.98.063845}
}

@article{shen2016experimental,
  title={Experimental realization of optomechanically induced non-reciprocity},
  author={Shen, Zhen and Zhang, Yan-Lei and Chen, Yuan and Zou, Chang-Ling and Xiao, Yun-Feng and Zou, Xu-Bo and Sun, Fang-Wen and Guo, Guang-Can and Dong, Chun-Hua},
  journal={Nat. Photonics},
  volume={10},
  number={10},
  pages={657--661},
  year={2016},
  publisher={Nature Publishing Group UK London},
  doi = {10.1038/nphoton.2016.161},
  url = {https://doi.org/10.1038/nphoton.2016.161}
}

@article{shen2018reconfigurable,
  title={Reconfigurable optomechanical circulator and directional amplifier},
  author={Shen, Zhen and Zhang, Yan-Lei and Chen, Yuan and Sun, Fang-Wen and Zou, Xu-Bo and Guo, Guang-Can and Zou, Chang-Ling and Dong, Chun-Hua},
  journal={Nat. Commun.},
  volume={9},
  number={1},
  pages={1797},
  year={2018},
  publisher={Nature Publishing Group},
  doi = {10.1038/s41467-018-04187-8},
  url = {https://doi.org/10.1038/s41467-018-04187-8}
}

@article{ZARERAMESHTI20221,
title = {Cavity magnonics},
journal = {Phys. Rep.},
volume = {979},
pages = {1-61},
year = {2022},
issn = {0370-1573},
doi = {https://doi.org/10.1016/j.physrep.2022.06.001},
url = {https://www.sciencedirect.com/science/article/pii/S0370157322002460},
author = {Babak {Zare Rameshti} and Silvia {Viola Kusminskiy} and James A. Haigh and Koji Usami and Dany Lachance-Quirion and Yasunobu Nakamura and Can-Ming Hu and Hong X. Tang and Gerrit E.W. Bauer and Yaroslav M. Blanter},
}

@article{YUAN20221,
title = {Quantum magnonics: When magnon spintronics meets quantum information science},
journal = {Phys. Rep.},
volume = {965},
pages = {1-74},
year = {2022},
issn = {0370-1573},
doi = {https://doi.org/10.1016/j.physrep.2022.03.002},
url = {https://www.sciencedirect.com/science/article/pii/S0370157322000977},
author = {H.Y. Yuan and Yunshan Cao and Akashdeep Kamra and Rembert A. Duine and Peng Yan},
}

@article{collet2016generation,
  title={Generation of coherent spin-wave modes in yttrium iron garnet microdiscs by spin--orbit torque},
  author={Collet, Martin and De Milly, Xavier and d’Allivy Kelly, Olivier and Naletov, Vladimir V and Bernard, Rozenn and Bortolotti, Paolo and Ben Youssef, J and Demidov, VE and Demokritov, SO and Prieto, Jose Luis and others},
  journal={Nat. Commun.},
  volume={7},
  number={1},
  pages={10377},
  year={2016},
  publisher={Nature Publishing Group UK London},
  doi = {10.1038/ncomms10377},
  url = {https://doi.org/10.1038/ncomms10377},
}

@article{PhysRevB.97.060405,
  title = {Thermodynamic entanglement of magnonic condensates},
  author = {Yuan, H. Y. and Yung, Man-Hong},
  journal = {Phys. Rev. B},
  volume = {97},
  issue = {6},
  pages = {060405},
  numpages = {6},
  year = {2018},
  month = {Feb},
  publisher = {American Physical Society},
  doi = {10.1103/PhysRevB.97.060405},
  url = {https://link.aps.org/doi/10.1103/PhysRevB.97.060405}
}

@article{PhysRevA.71.032317,
  title = {Macroscopic entanglement of many-magnon states},
  author = {Morimae, Tomoyuki and Sugita, Ayumu and Shimizu, Akira},
  journal = {Phys. Rev. A},
  volume = {71},
  issue = {3},
  pages = {032317},
  numpages = {12},
  year = {2005},
  month = {Mar},
  publisher = {American Physical Society},
  doi = {10.1103/PhysRevA.71.032317},
  url = {https://link.aps.org/doi/10.1103/PhysRevA.71.032317}
}

@article{zhang2015cavity,
  title={Cavity quantum electrodynamics with ferromagnetic magnons in a small yttrium-iron-garnet sphere},
  author={Zhang, Dengke and Wang, Xin-Ming and Li, Tie-Fu and Luo, Xiao-Qing and Wu, Weidong and Nori, Franco and You, JQ},
  journal={npj Quantum Information},
  volume={1},
  number={1},
  pages={1--6},
  year={2015},
  publisher={Nature Publishing Group},
  doi = {10.1038/npjqi.2015.14},
  url = {https://doi.org/10.1038/npjqi.2015.14}
}

@article{PhysRevB.94.224410,
  title = {Magnon Kerr effect in a strongly coupled cavity-magnon system},
  author = {Wang, Yi-Pu and Zhang, Guo-Qiang and Zhang, Dengke and Luo, Xiao-Qing and Xiong, Wei and Wang, Shuai-Peng and Li, Tie-Fu and Hu, C.-M. and You, J. Q.},
  journal = {Phys. Rev. B},
  volume = {94},
  issue = {22},
  pages = {224410},
  numpages = {8},
  year = {2016},
  month = {Dec},
  publisher = {American Physical Society},
  doi = {10.1103/PhysRevB.94.224410},
  url = {https://link.aps.org/doi/10.1103/PhysRevB.94.224410}
}

@article{zhang2019theory,
  title={Theory of the magnon Kerr effect in cavity magnonics},
  author={Zhang, GuoQiang and Wang, YiPu and You, JianQiang},
  journal={Sci. China Phys. Mech. Astron.},
  volume={62},
  pages={987511},
  year={2019},
  publisher={Springer},
  doi = {10.1007/s11433-018-9344-8},
  url = {https://doi.org/10.1007/s11433-018-9344-8}
}

@article{PhysRevLett.120.057202,
  title = {Bistability of Cavity Magnon Polaritons},
  author = {Wang, Yi-Pu and Zhang, Guo-Qiang and Zhang, Dengke and Li, Tie-Fu and Hu, C.-M. and You, J. Q.},
  journal = {Phys. Rev. Lett.},
  volume = {120},
  issue = {5},
  pages = {057202},
  numpages = {6},
  year = {2018},
  month = {Jan},
  publisher = {American Physical Society},
  doi = {10.1103/PhysRevLett.120.057202},
  url = {https://link.aps.org/doi/10.1103/PhysRevLett.120.057202}
}

@article{zhang2017observation,
  title={Observation of the exceptional point in cavity magnon-polaritons},
  author={Zhang, Dengke and Luo, Xiao-Qing and Wang, Yi-Pu and Li, Tie-Fu and You, JQ},
  journal={Nat. Commun.},
  volume={8},
  number={1},
  pages={1368},
  year={2017},
  publisher={Nature Publishing Group UK London},
  doi = {10.1038/s41467-017-01634-w},
  url = {https://doi.org/10.1038/s41467-017-01634-w}
}

@article{PhysRevB.91.094423,
  title = {Exchange magnon-polaritons in microwave cavities},
  author = {Cao, Yunshan and Yan, Peng and Huebl, Hans and Goennenwein, Sebastian T. B. and Bauer, Gerrit E. W.},
  journal = {Phys. Rev. B},
  volume = {91},
  issue = {9},
  pages = {094423},
  numpages = {6},
  year = {2015},
  month = {Mar},
  publisher = {American Physical Society},
  doi = {10.1103/PhysRevB.91.094423},
  url = {https://link.aps.org/doi/10.1103/PhysRevB.91.094423}
}

@article{PhysRevLett.118.217201,
  title = {Cavity Mediated Manipulation of Distant Spin Currents Using a Cavity-Magnon-Polariton},
  author = {Bai, Lihui and Harder, Michael and Hyde, Paul and Zhang, Zhaohui and Hu, Can-Ming and Chen, Y. P. and Xiao, John Q.},
  journal = {Phys. Rev. Lett.},
  volume = {118},
  issue = {21},
  pages = {217201},
  numpages = {5},
  year = {2017},
  month = {May},
  publisher = {American Physical Society},
  doi = {10.1103/PhysRevLett.118.217201},
  url = {https://link.aps.org/doi/10.1103/PhysRevLett.118.217201}
}

@article{PhysRevB.98.174423,
  title = {Direct measurement of foldover in cavity magnon-polariton systems},
  author = {Hyde, P. and Yao, B. M. and Gui, Y. S. and Zhang, Guo-Qiang and You, J. Q. and Hu, C.-M.},
  journal = {Phys. Rev. B},
  volume = {98},
  issue = {17},
  pages = {174423},
  numpages = {9},
  year = {2018},
  month = {Nov},
  publisher = {American Physical Society},
  doi = {10.1103/PhysRevB.98.174423},
  url = {https://link.aps.org/doi/10.1103/PhysRevB.98.174423}
}

@article{PhysRevB.99.054404,
  title = {Higher-order exceptional point in a cavity magnonics system},
  author = {Zhang, Guo-Qiang and You, J. Q.},
  journal = {Phys. Rev. B},
  volume = {99},
  issue = {5},
  pages = {054404},
  numpages = {9},
  year = {2019},
  month = {Feb},
  publisher = {American Physical Society},
  doi = {10.1103/PhysRevB.99.054404},
  url = {https://link.aps.org/doi/10.1103/PhysRevB.99.054404}
}

@article{PhysRevA.103.063708,
  title = {Exceptional-point-engineered cavity magnomechanics},
  author = {Lu, Tian-Xiang and Zhang, Huilai and Zhang, Qian and Jing, Hui},
  journal = {Phys. Rev. A},
  volume = {103},
  issue = {6},
  pages = {063708},
  numpages = {9},
  year = {2021},
  month = {Jun},
  publisher = {American Physical Society},
  doi = {10.1103/PhysRevA.103.063708},
  url = {https://link.aps.org/doi/10.1103/PhysRevA.103.063708}
}

@article{Zhang:25,
author = {Haoran Zhang and Chao Wang and Qiwen Bao and Zhenzhen Liu and Haoliang Liu and Jun-Jun Xiao},
journal = {Opt. Lett.},
number = {13},
pages = {4154--4157},
publisher = {Optica Publishing Group},
title = {Topological mode switch by exceptional point encircling in a cavity magnonic system},
volume = {50},
month = {Jul},
year = {2025},
url = {https://opg.optica.org/ol/abstract.cfm?URI=ol-50-13-4154},
doi = {10.1364/OL.567749},
}

@article{PhysRevLett.113.083603,
  title = {Hybridizing Ferromagnetic Magnons and Microwave Photons in the Quantum Limit},
  author = {Tabuchi, Yutaka and Ishino, Seiichiro and Ishikawa, Toyofumi and Yamazaki, Rekishu and Usami, Koji and Nakamura, Yasunobu},
  journal = {Phys. Rev. Lett.},
  volume = {113},
  issue = {8},
  pages = {083603},
  numpages = {5},
  year = {2014},
  month = {Aug},
  publisher = {American Physical Society},
  doi = {10.1103/PhysRevLett.113.083603},
  url = {https://link.aps.org/doi/10.1103/PhysRevLett.113.083603}
}

@article{
doi:10.1126/science.aaa3693,
author = {Yutaka Tabuchi  and Seiichiro Ishino  and Atsushi Noguchi  and Toyofumi Ishikawa  and Rekishu Yamazaki  and Koji Usami  and Yasunobu Nakamura },
title = {Coherent coupling between a ferromagnetic magnon and a superconducting qubit},
journal = {Science},
volume = {349},
number = {6246},
pages = {405-408},
year = {2015},
doi = {10.1126/science.aaa3693},
URL = {https://www.science.org/doi/abs/10.1126/science.aaa3693},
}

@article{PhysRevLett.104.077202,
  title = {Strong Field Interactions between a Nanomagnet and a Photonic Cavity},
  author = {Soykal, \"O. O. and Flatt\'e, M. E.},
  journal = {Phys. Rev. Lett.},
  volume = {104},
  issue = {7},
  pages = {077202},
  numpages = {4},
  year = {2010},
  month = {Feb},
  publisher = {American Physical Society},
  doi = {10.1103/PhysRevLett.104.077202},
  url = {https://link.aps.org/doi/10.1103/PhysRevLett.104.077202}
}

@article{PhysRevLett.113.156401,
  title = {Strongly Coupled Magnons and Cavity Microwave Photons},
  author = {Zhang, Xufeng and Zou, Chang-Ling and Jiang, Liang and Tang, Hong X.},
  journal = {Phys. Rev. Lett.},
  volume = {113},
  issue = {15},
  pages = {156401},
  numpages = {5},
  year = {2014},
  month = {Oct},
  publisher = {American Physical Society},
  doi = {10.1103/PhysRevLett.113.156401},
  url = {https://link.aps.org/doi/10.1103/PhysRevLett.113.156401}
}

@article{PhysRevB.93.144420,
  title = {Ultrahigh cooperativity interactions between magnons and resonant photons in a YIG sphere},
  author = {Bourhill, J. and Kostylev, N. and Goryachev, M. and Creedon, D. L. and Tobar, M. E.},
  journal = {Phys. Rev. B},
  volume = {93},
  issue = {14},
  pages = {144420},
  numpages = {8},
  year = {2016},
  month = {Apr},
  publisher = {American Physical Society},
  doi = {10.1103/PhysRevB.93.144420},
  url = {https://link.aps.org/doi/10.1103/PhysRevB.93.144420}
}

@article{PhysRevLett.121.137203,
  title = {Level Attraction Due to Dissipative Magnon-Photon Coupling},
  author = {Harder, M. and Yang, Y. and Yao, B. M. and Yu, C. H. and Rao, J. W. and Gui, Y. S. and Stamps, R. L. and Hu, C.-M.},
  journal = {Phys. Rev. Lett.},
  volume = {121},
  issue = {13},
  pages = {137203},
  numpages = {5},
  year = {2018},
  month = {Sep},
  publisher = {American Physical Society},
  doi = {10.1103/PhysRevLett.121.137203},
  url = {https://link.aps.org/doi/10.1103/PhysRevLett.121.137203}
}

@article{yao2017cooperative,
  title={Cooperative polariton dynamics in feedback-coupled cavities},
  author={Yao, Bimu and Gui, YS and Rao, JW and Kaur, S and Chen, XS and Lu, W and Xiao, Y and Guo, H and Marzlin, K-P and Hu, C-M},
  journal={Nat. Commun.},
  volume={8},
  number={1},
  pages={1437},
  year={2017},
  publisher={Nature Publishing Group UK London},
  doi = {10.1038/s41467-017-01796-7},
  url = {https://doi.org/10.1038/s41467-017-01796-7}
}

@article{Lachance-Quirion_2019,
doi = {10.7567/1882-0786/ab248d},
url = {https://dx.doi.org/10.7567/1882-0786/ab248d},
year = {2019},
month = {jun},
publisher = {IOP Publishing},
volume = {12},
number = {7},
pages = {070101},
author = {Lachance-Quirion, Dany and Tabuchi, Yutaka and Gloppe, Arnaud and Usami, Koji and Nakamura, Yasunobu},
title = {Hybrid quantum systems based on magnonics},
journal = {Appl. Phys. Express},
}

@article{PhysRevB.109.L041301,
  title = {Quantum-enhanced metrology in cavity magnonics},
  author = {Wan, Qing-Kun and Shi, Hai-Long and Guan, Xi-Wen},
  journal = {Phys. Rev. B},
  volume = {109},
  issue = {4},
  pages = {L041301},
  numpages = {8},
  year = {2024},
  month = {Jan},
  publisher = {American Physical Society},
  doi = {10.1103/PhysRevB.109.L041301},
  url = {https://link.aps.org/doi/10.1103/PhysRevB.109.L041301}
}

@article{Flower_2019,
doi = {10.1088/1367-2630/ab3e1c},
url = {https://dx.doi.org/10.1088/1367-2630/ab3e1c},
year = {2019},
month = {sep},
publisher = {IOP Publishing},
volume = {21},
number = {9},
pages = {095004},
author = {Flower, Graeme and Goryachev, Maxim and Bourhill, Jeremy and Tobar, Michael E},
title = {Experimental implementations of cavity-magnon systems: from ultra strong coupling to applications in precision measurement},
journal = {New J. Phys.},
}

@article{PhysRevA.103.052419,
  title = {Microwave quantum illumination via cavity magnonics},
  author = {Cai, Qizhi and Liao, Jinkun and Shen, Bohai and Guo, Guangcan and Zhou, Qiang},
  journal = {Phys. Rev. A},
  volume = {103},
  issue = {5},
  pages = {052419},
  numpages = {10},
  year = {2021},
  month = {May},
  publisher = {American Physical Society},
  doi = {10.1103/PhysRevA.103.052419},
  url = {https://link.aps.org/doi/10.1103/PhysRevA.103.052419}
}

@article{PhysRevLett.117.133602,
  title = {Triple-Resonant Brillouin Light Scattering in Magneto-Optical Cavities},
  author = {Haigh, J. A. and Nunnenkamp, A. and Ramsay, A. J. and Ferguson, A. J.},
  journal = {Phys. Rev. Lett.},
  volume = {117},
  issue = {13},
  pages = {133602},
  numpages = {6},
  year = {2016},
  month = {Sep},
  publisher = {American Physical Society},
  doi = {10.1103/PhysRevLett.117.133602},
  url = {https://link.aps.org/doi/10.1103/PhysRevLett.117.133602}
}

@article{PhysRevB.93.174427,
  title = {Bidirectional conversion between microwave and light via ferromagnetic magnons},
  author = {Hisatomi, R. and Osada, A. and Tabuchi, Y. and Ishikawa, T. and Noguchi, A. and Yamazaki, R. and Usami, K. and Nakamura, Y.},
  journal = {Phys. Rev. B},
  volume = {93},
  issue = {17},
  pages = {174427},
  numpages = {13},
  year = {2016},
  month = {May},
  publisher = {American Physical Society},
  doi = {10.1103/PhysRevB.93.174427},
  url = {https://link.aps.org/doi/10.1103/PhysRevB.93.174427}
}

@article{zhang2015magnon,
  title={Magnon dark modes and gradient memory},
  author={Zhang, Xufeng and Zou, Chang-Ling and Zhu, Na and Marquardt, Florian and Jiang, Liang and Tang, Hong X},
  journal={Nat. Commun.},
  volume={6},
  number={1},
  pages={8914},
  year={2015},
  publisher={Nature Publishing Group UK London},
  doi = {10.1038/ncomms9914},
  url = {https://doi.org/10.1038/ncomms9914}
}

@article{PhysRevLett.127.183202,
  title = {Long-Time Memory and Ternary Logic Gate Using a Multistable Cavity Magnonic System},
  author = {Shen, Rui-Chang and Wang, Yi-Pu and Li, Jie and Zhu, Shi-Yao and Agarwal, G. S. and You, J. Q.},
  journal = {Phys. Rev. Lett.},
  volume = {127},
  issue = {18},
  pages = {183202},
  numpages = {6},
  year = {2021},
  month = {Oct},
  publisher = {American Physical Society},
  doi = {10.1103/PhysRevLett.127.183202},
  url = {https://link.aps.org/doi/10.1103/PhysRevLett.127.183202}
}

@article{PhysRevLett.116.223601,
  title = {Cavity Optomagnonics with Spin-Orbit Coupled Photons},
  author = {Osada, A. and Hisatomi, R. and Noguchi, A. and Tabuchi, Y. and Yamazaki, R. and Usami, K. and Sadgrove, M. and Yalla, R. and Nomura, M. and Nakamura, Y.},
  journal = {Phys. Rev. Lett.},
  volume = {116},
  issue = {22},
  pages = {223601},
  numpages = {5},
  year = {2016},
  month = {Jun},
  publisher = {American Physical Society},
  doi = {10.1103/PhysRevLett.116.223601},
  url = {https://link.aps.org/doi/10.1103/PhysRevLett.116.223601}
}

@article{PhysRevApplied.12.034001,
  title = {Magnon-Induced Nonreciprocity Based on the Magnon Kerr Effect},
  author = {Kong, Cui and Xiong, Hao and Wu, Ying},
  journal = {Phys. Rev. Appl.},
  volume = {12},
  issue = {3},
  pages = {034001},
  numpages = {9},
  year = {2019},
  month = {Sep},
  publisher = {American Physical Society},
  doi = {10.1103/PhysRevApplied.12.034001},
  url = {https://link.aps.org/doi/10.1103/PhysRevApplied.12.034001}
}

@article{Ren:21,
author = {Ya-Long Ren and Sheng-Li Ma and Ji-Kun Xie and Xin-Ke Li and Fu-Li Li},
journal = {Opt. Express},
number = {25},
pages = {41399--41408},
publisher = {Optica Publishing Group},
title = {Non-reciprocal optical transmission in cavity optomagnonics},
volume = {29},
month = {Dec},
year = {2021},
url = {https://opg.optica.org/oe/abstract.cfm?URI=oe-29-25-41399},
doi = {10.1364/OE.440697},
}

@article{Kong:21,
author = {Cui Kong and Xi-Min Bao and Ji-Bing Liu and Hao Xiong},
journal = {Opt. Express},
number = {16},
pages = {25477--25487},
publisher = {Optica Publishing Group},
title = {Magnon-mediated nonreciprocal microwave transmission based on quantum interference},
volume = {29},
month = {Aug},
year = {2021},
url = {https://opg.optica.org/oe/abstract.cfm?URI=oe-29-16-25477},
doi = {10.1364/OE.430619},
}

@article{PhysRevLett.120.133602,
  title = {Brillouin Light Scattering by Magnetic Quasivortices in Cavity Optomagnonics},
  author = {Osada, A. and Gloppe, A. and Hisatomi, R. and Noguchi, A. and Yamazaki, R. and Nomura, M. and Nakamura, Y. and Usami, K.},
  journal = {Phys. Rev. Lett.},
  volume = {120},
  issue = {13},
  pages = {133602},
  numpages = {5},
  year = {2018},
  month = {Mar},
  publisher = {American Physical Society},
  doi = {10.1103/PhysRevLett.120.133602},
  url = {https://link.aps.org/doi/10.1103/PhysRevLett.120.133602}
}

@article{PhysRevLett.117.123605,
  title = {Optomagnonic Whispering Gallery Microresonators},
  author = {Zhang, Xufeng and Zhu, Na and Zou, Chang-Ling and Tang, Hong X.},
  journal = {Phys. Rev. Lett.},
  volume = {117},
  issue = {12},
  pages = {123605},
  numpages = {5},
  year = {2016},
  month = {Sep},
  publisher = {American Physical Society},
  doi = {10.1103/PhysRevLett.117.123605},
  url = {https://link.aps.org/doi/10.1103/PhysRevLett.117.123605}
}

@article{PhysRevLett.123.127202,
  title = {Nonreciprocity and Unidirectional Invisibility in Cavity Magnonics},
  author = {Wang, Yi-Pu and Rao, J. W. and Yang, Y. and Xu, Peng-Chao and Gui, Y. S. and Yao, B. M. and You, J. Q. and Hu, C.-M.},
  journal = {Phys. Rev. Lett.},
  volume = {123},
  issue = {12},
  pages = {127202},
  numpages = {6},
  year = {2019},
  month = {Sep},
  publisher = {American Physical Society},
  doi = {10.1103/PhysRevLett.123.127202},
  url = {https://link.aps.org/doi/10.1103/PhysRevLett.123.127202}
}

@article{MING2024107915,
title = {Nonreciprocal transmission, reflection, and absorption in non-Hermitian cavity magnonics},
journal = {Results. Phys.},
volume = {64},
pages = {107915},
year = {2024},
issn = {2211-3797},
doi = {https://doi.org/10.1016/j.rinp.2024.107915},
url = {https://www.sciencedirect.com/science/article/pii/S2211379724006004},
author = {Ying Ming and Rong-Can Yang},
}

@article{PhysRevA.110.043704,
  title = {Nonreciprocal quantum phase transition in cavity magnonics},
  author = {Xu, Ye-Jun and Zhai, Long-Hua and Fu, Peng and Cheng, Shou-Jing and Zhang, Guo-Qiang},
  journal = {Phys. Rev. A},
  volume = {110},
  issue = {4},
  pages = {043704},
  numpages = {10},
  year = {2024},
  month = {Oct},
  publisher = {American Physical Society},
  doi = {10.1103/PhysRevA.110.043704},
  url = {https://link.aps.org/doi/10.1103/PhysRevA.110.043704}
}

@article{Xie:23,
author = {Hong Xie and Le-Wei He and Chang-Geng Liao and Zhi-Hua Chen and Xiu-Min Lin},
journal = {Opt. Express},
number = {5},
pages = {7994--8004},
publisher = {Optica Publishing Group},
title = {Generation of robust optical entanglement in cavity optomagnonics},
volume = {31},
month = {Feb},
year = {2023},
url = {https://opg.optica.org/oe/abstract.cfm?URI=oe-31-5-7994},
doi = {10.1364/OE.478963},
}

@article{Maayani2018,
  title={Flying couplers above spinning resonators generate irreversible refraction},
  author={Maayani, S. and Dahan, R. and Kligerman, Y. and Moses, E. and Hassan, A. U. and Jing, H. and Nori, F. and Christodoulides, D. N. and Carmon, T.},
  journal={Nature},
  volume={558},
  issue = {7711},
  pages={569--572},
  year={2018},
  doi = {10.1038/s41586-018-0245-5},
  url = {https://doi.org/10.1038/s41586-018-0245-5}
}

@article{RevModPhys.86.1391,
  title = {Cavity optomechanics},
  author = {Aspelmeyer, Markus and Kippenberg, Tobias J. and Marquardt, Florian},
  journal = {Rev. Mod. Phys.},
  volume = {86},
  issue = {4},
  pages = {1391--1452},
  numpages = {62},
  year = {2014},
  month = {Dec},
  publisher = {American Physical Society},
  doi = {10.1103/RevModPhys.86.1391},
  url = {https://link.aps.org/doi/10.1103/RevModPhys.86.1391}
}

@article{PhysRevA.99.021801,
  title = {Squeezed states of magnons and phonons in cavity magnomechanics},
  author = {Li, Jie and Zhu, Shi-Yao and Agarwal, G. S.},
  journal = {Phys. Rev. A},
  volume = {99},
  issue = {2},
  pages = {021801},
  numpages = {6},
  year = {2019},
  month = {Feb},
  publisher = {American Physical Society},
  doi = {10.1103/PhysRevA.99.021801},
  url = {https://link.aps.org/doi/10.1103/PhysRevA.99.021801}
}

@article{PhysRevA.108.063703,
  title = {Magnon squeezing by two-tone driving of a qubit in cavity-magnon-qubit systems},
  author = {Guo, Qi and Cheng, Jiong and Tan, Huatang and Li, Jie},
  journal = {Phys. Rev. A},
  volume = {108},
  issue = {6},
  pages = {063703},
  numpages = {7},
  year = {2023},
  month = {Dec},
  publisher = {American Physical Society},
  doi = {10.1103/PhysRevA.108.063703},
  url = {https://link.aps.org/doi/10.1103/PhysRevA.108.063703}
}

@article{Ding:22,
author = {Ming-Song Ding and Li Zheng and Ying Shi and Yu-Jie Liu},
journal = {J. Opt. Soc. Am. B},
number = {10},
pages = {2665--2669},
publisher = {Optica Publishing Group},
title = {Magnon squeezing enhanced entanglement in a cavity magnomechanical system},
volume = {39},
month = {Oct},
year = {2022},
url = {https://opg.optica.org/josab/abstract.cfm?URI=josab-39-10-2665},
doi = {10.1364/JOSAB.465554}
}

@article{PhysRevA.111.053707,
  title = {Magnon squeezing induced by virtual photons in a magnon-cavity-qubit system},
  author = {Xia, Ai-Bo and Cheng, Jing and Tian, Ding-Li and Han, Cheng-Jun and Wang, Y. P.},
  journal = {Phys. Rev. A},
  volume = {111},
  issue = {5},
  pages = {053707},
  numpages = {9},
  year = {2025},
  month = {May},
  publisher = {American Physical Society},
  doi = {10.1103/PhysRevA.111.053707},
  url = {https://link.aps.org/doi/10.1103/PhysRevA.111.053707}
}

@article{PhysRevA.100.062501,
  title = {Emission of photon pairs by mechanical stimulation of the squeezed vacuum},
  author = {Qin, Wei and Macr\`{\i}, Vincenzo and Miranowicz, Adam and Savasta, Salvatore and Nori, Franco},
  journal = {Phys. Rev. A},
  volume = {100},
  issue = {6},
  pages = {062501},
  numpages = {17},
  year = {2019},
  month = {Dec},
  publisher = {American Physical Society},
  doi = {10.1103/PhysRevA.100.062501},
  url = {https://link.aps.org/doi/10.1103/PhysRevA.100.062501}
}

@article{PhysRevResearch.1.023021,
  title = {Quantum entanglement between two magnon modes via Kerr nonlinearity driven far from equilibrium},
  author = {Zhang, Zhedong and Scully, Marlan O. and Agarwal, Girish S.},
  journal = {Phys. Rev. Res.},
  volume = {1},
  issue = {2},
  pages = {023021},
  numpages = {7},
  year = {2019},
  month = {Sep},
  publisher = {American Physical Society},
  doi = {10.1103/PhysRevResearch.1.023021},
  url = {https://link.aps.org/doi/10.1103/PhysRevResearch.1.023021}
}

@article{righini2011whispering,
  title={Whispering gallery mode microresonators: fundamentals and applications},
  author={Righini, Giancarlo C and Dumeige, Yannick and F{\'e}ron, Patrice and Ferrari, Maurizio and Nunzi Conti, Gualtiero and Ristic, Davor and Soria, Silvia},
  journal={Riv. Nuovo Cim.},
  volume={34},
  pages={435--488},
  year={2011},
  publisher={Springer},
  doi = {10.1393/ncr/i2011-10067-2},
  url = {https://doi.org/10.1393/ncr/i2011-10067-2}
}

@article{
doi:10.1126/sciadv.1501286,
author = {Xufeng Zhang  and Chang-Ling Zou  and Liang Jiang  and Hong X. Tang },
title = {Cavity magnomechanics},
journal = {Sci. Adv.},
volume = {2},
number = {3},
pages = {e1501286},
year = {2016},
doi = {10.1126/sciadv.1501286},
URL = {https://www.science.org/doi/abs/10.1126/sciadv.1501286},
}

@article{PhysRevLett.114.080503,
  title = {Microwave Quantum Illumination},
  author = {Barzanjeh, Shabir and Guha, Saikat and Weedbrook, Christian and Vitali, David and Shapiro, Jeffrey H. and Pirandola, Stefano},
  journal = {Phys. Rev. Lett.},
  volume = {114},
  issue = {8},
  pages = {080503},
  numpages = {5},
  year = {2015},
  month = {Feb},
  publisher = {American Physical Society},
  doi = {10.1103/PhysRevLett.114.080503},
  url = {https://link.aps.org/doi/10.1103/PhysRevLett.114.080503}
}

@article{PhysRevLett.134.196904,
  title = {Nonreciprocal Control of the Speed of Light Using Cavity Magnonics},
  author = {Yao, Jiguang and Lu, Chenyang and Fan, Xiaolong and Xue, Desheng and Bridges, Greg E. and Hu, C.-M.},
  journal = {Phys. Rev. Lett.},
  volume = {134},
  issue = {19},
  pages = {196904},
  numpages = {7},
  year = {2025},
  month = {May},
  publisher = {American Physical Society},
  doi = {10.1103/PhysRevLett.134.196904},
  url = {https://link.aps.org/doi/10.1103/PhysRevLett.134.196904}
}

@article{PhysRevResearch.3.023126,
  title = {Bistability of squeezing and entanglement in cavity magnonics},
  author = {Yang, Zhi-Bo and Jin, Hua and Jin, Jing-Wen and Liu, Jian-Yu and Liu, Hong-Yu and Yang, Rong-Can},
  journal = {Phys. Rev. Res.},
  volume = {3},
  issue = {2},
  pages = {023126},
  numpages = {9},
  year = {2021},
  month = {May},
  publisher = {American Physical Society},
  doi = {10.1103/PhysRevResearch.3.023126},
  url = {https://link.aps.org/doi/10.1103/PhysRevResearch.3.023126}
}

@article{Wang_2025,
doi = {10.1088/1751-8121/adde01},
url = {https://dx.doi.org/10.1088/1751-8121/adde01},
year = {2025},
month = {jun},
publisher = {IOP Publishing},
volume = {58},
number = {23},
pages = {235302},
author = {Wang, Jia-Xin and Guo, Qi and Zhang, Yuchi and Li, Gang and Zhang, Tiancai},
title = {Steady-squeezed-state generation via Kerr nonlinearity in cavity magnonics system},
journal = {J. Phys. Math. Theor.},
}

@article{Mathkoor2025,
  title={Bipartite entanglement and Gaussian quantum steering in a whispering gallery mode coupled with two magnon modes},
  author={Mathkoor, Faisal HA and Singh, SK and Ahmed, Rizwan and Peng, Jia-Xin and Amazioug, M and Khalid, Mohammad and Sohail, Amjad},
  journal={Sci. Rep.},
  volume={15},
  number={1},
  pages={13503},
  year={2025},
  publisher={Nature Publishing Group UK London},
doi = {10.1038/s41598-025-98011-1},
  url = {https://doi.org/10.1038/s41598-025-98011-1}
}

@article{PhysRevLett.121.087205,
  title = {Optical Cooling of Magnons},
  author = {Sharma, Sanchar and Blanter, Yaroslav M. and Bauer, Gerrit E. W.},
  journal = {Phys. Rev. Lett.},
  volume = {121},
  issue = {8},
  pages = {087205},
  numpages = {6},
  year = {2018},
  month = {Aug},
  publisher = {American Physical Society},
  doi = {10.1103/PhysRevLett.121.087205},
  url = {https://link.aps.org/doi/10.1103/PhysRevLett.121.087205}
}

@article{PhysRevA.109.043512,
  title = {Nonreciprocal photon-phonon entanglement in Kerr-modified spinning cavity magnomechanics},
  author = {Chen, Jiaojiao and Fan, Xiao-Gang and Xiong, Wei and Wang, Dong and Ye, Liu},
  journal = {Phys. Rev. A},
  volume = {109},
  issue = {4},
  pages = {043512},
  numpages = {9},
  year = {2024},
  month = {Apr},
  publisher = {American Physical Society},
  doi = {10.1103/PhysRevA.109.043512},
  url = {https://link.aps.org/doi/10.1103/PhysRevA.109.043512}
}

@article{10.1063/5.0190162,
    author = {Zheng, Qianjun and Zhong, Wenxue and Cheng, Guangling and Chen, Aixi},
    title = {Nonreciprocal microwave-optical entanglement in a magnon-based hybrid system},
    journal = {J. Appl. Phys.},
    volume = {135},
    number = {8},
    pages = {084401},
    year = {2024},
    month = {02},
    issn = {0021-8979},
    doi = {10.1063/5.0190162},
    url = {https://doi.org/10.1063/5.0190162}
}
\end{document}